\documentclass[aps,twocolumn,amsmath,amssymb,prb,superscriptaddress]{revtex4-1}
\usepackage{graphicx}
\usepackage{dcolumn}
\usepackage{bm}
\usepackage{color}	
\usepackage{ulem}	
\usepackage{xparse}
\usepackage{physics}
\usepackage{mathtools}
\usepackage{mathrsfs}
\usepackage{float}
\usepackage{afterpage}
\usepackage[ppl]{mathcomp}
\usepackage{autobreak}

\begin{document}

\preprint{APS/PRB}


\title{Magnetic Anisotropies and Skyrmion Lattice Related to Magnetic Quadrupole Interactions 
of the RKKY Mechanism in Frustrated 
Spin-Trimer System Gd$_{3}$Ru$_{4}$Al$_{12}$ with a Breathing Kagome Structure
}

\date{\today}

\author{S. Nakamura}
\thanks{corresponding authors}
\affiliation{Institute for Materials Research, Tohoku University, Katahira 2--1--1, Sendai 980-8577, Japan}
\affiliation{Center for Low Temperature Science, Tohoku University, Katahira 2--1--1, Sendai 980-8577, Japan}

\begin{abstract}
{The origin of the magnetic quadrupole (MQ) interactions in Gd$_{3}$Ru$_{4}$Al$_{12}$ which is known
as a frustrated spin system and as a host material of skyrmion with a sentrosymmetric crystal structure are discussed.
The MQ interactions between ferromagnetic (FM) spin trimers with imperfect FM directivity are deduced from synthesis 
of dipole Ruderman-Kittel-Kasuya-Yosida (RKKY) interactions. 
The Hamiltonian which includes both the MQ interactions and dipole interactions is proposed, and magnetic anisotropies, 
magnetic phase transitions and contribution of the MQ interactions to stabilize the skyrmion lattice (SkL) are discussed
based on this Hamiltonian. 
Degrees of MQ freedom carried by the trimers contribute to stabilizing SkL which appears at finite temperatures.
}
\end{abstract}


\maketitle

\section{Introduction}

The 4$f$ electrons in rare earth compounds locate in the xenon-closed-shells and their spacial extents are 
smaller than those of 3$d$ electrons in transition element compounds
\cite{Kasuya1987}.  
This small spacial extents of 4$f$ electrons results in weak direct exchange interactions between 4$f$ electrons. 
In metallic 4$f$ electron systems, the RKKY mechanism which is mediated by conduction electrons play key roles to
generate interactions between 4$f$ electrons even if the concentrations of magnetic ions are high.
The weak direct interactions leads to the characteristic magnetism of metallic rare earth compounds,
such as the dense Kondo effect and heavy fermion states
\cite{Stewart1984}. 
When we study metallic frustrated spin systems, rare earth compounds are fascinating research subjects
because there is some simplicity in theoretical treatment arising from the strong localization of 4$f$ electrons.
However, we have to take the characteristics of long reaching distance and 
vibrational behaviors of the RKKY interactions into consideration.

\begin{figure}[t]
		\includegraphics[width=7cm]{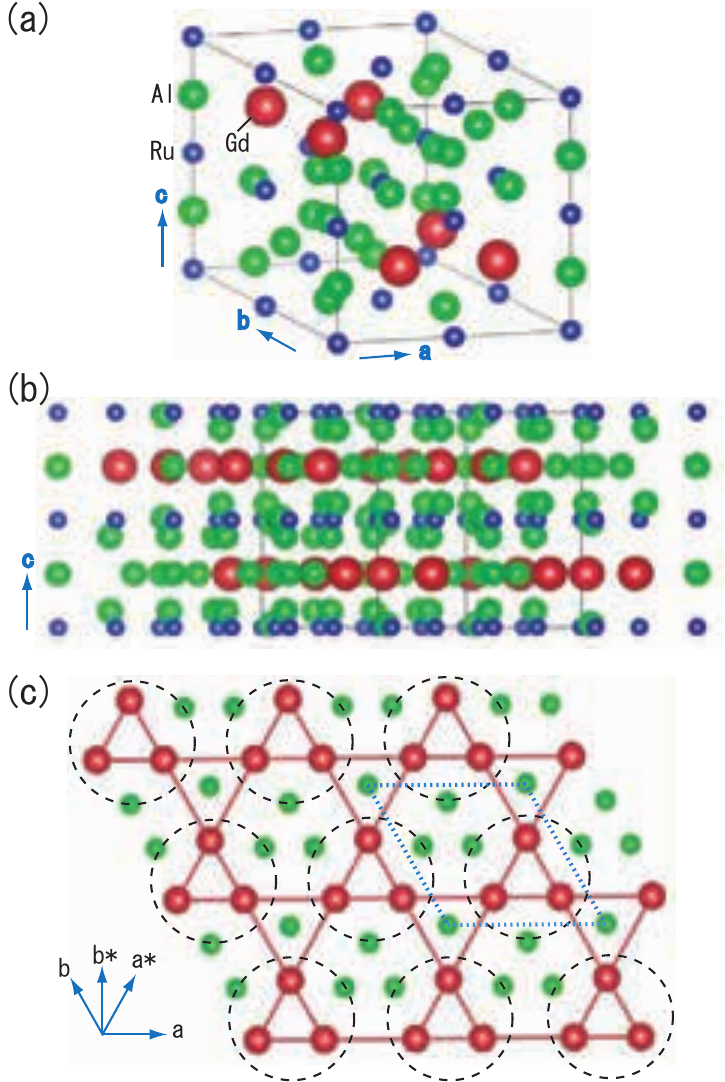}	
		\caption{(a) Crystal structure of Gd$_{3}$Ru$_{4}$Al$_{12}$
		\cite{Niermann2002}.
		Drawing of the crystal structure was produced using VESTA
		\cite{VESTA}.		
		The red (large), blue (small),
		and light green (medium-sized) spheres denote Gd, Ru, and Al ions, respectively. 
		The lattice constants are obtained to be 0.8778 nm for $a$ axis and 0.9472 nm for $c$ axis
\cite{Nakamura2018}
		(b) Structure projected parallel to the $ab$ plane. 
		(c) A Gd--Al layer projected parallel to the $c$ axis.
		Bonds are drawn between the nearest neighbor and the next nearest neighbor of the Gd ions. 
		The broken circles indicate FM trimers which form a triangular lattice.
		The arrow ${\bm a}^{*}$ and ${\bm b}^{*}$ denote
		the directions of the reciprocal lattice vectors of ${\bm a}$ and ${\bm b}$, respectively.
		The dotted-blue rhombus indicates an unit cell. 
		}
		\label{f1}
	\end{figure}

The ternary frustrated spin system Gd$_{3}$Ru$_{4}$Al$_{12}$ crystallize
in a hexagonal and centrosymmetric structure, which belongs to the space group $P6_3/ mmc$ (No. 194)
\cite{Niermann2002}. 
In this crystal, magnetic Gd--Al layers and nonmagnetic Ru--Al layers are stack 
alternately along the $c$ axis as illustrated in Fig.~\ref{f1} (a) and (b). 
In the Gd--Al layers, Gd$^{3+}$ ions ($S=7/2$, $L=0$) form a breathing kagome lattice
composed of two different sized regular triangles as shown in Fig.~\ref{f1} (c).
The lattice constants are obtained to be 0.8778 nm for $a$ axis and 0.9472 nm for $c$ axis by a x-ray diffraction method
\cite{Nakamura2018}. 
In the low temperature range, ferromagnetic (FM) spin trimers are generated on the
small sized Gd-triangles in Fig.~\ref{f1} (c) by the RKKY interaction
\cite{Nakamura2018}. 
The binding energy of this FM trimer is estimated to be 184 K.
In association with this formation of FM trimers, it is believed that the breathing kagome lattice of 
Gd$_{3}$Ru$_{4}$Al$_{12}$ effectively transforms into an antiferromagnetic (AFM) triangular lattice.
The broken circles in  Fig.~\ref{f1} (c) denotes the FM trimers, and these trimers form a triangular lattice in the Gd-Al plane.
As the temperature further decreases, Gd$_{3}$Ru$_{4}$Al$_{12}$ exhibits 
successive phase transitions at $T_{2}=18.6$ K and $T_{1}=17.5$ K.
The existence of the intermediate temperature (IMT) phase and reduced magnetic entropy (40\% of $R \ln 8$) at 
$T_{2}$ are indications of geometrical frustration.

Recently, anisotropic magnetic properties of Gd$_{3}$Ru$_{4}$Al$_{12}$ were reported, and it was pointed out
that easy axis and easy plane types of magnetic anisotropies coexist in this compound
\cite{Nakamura2023}. 
This unusual features are phenomenologically explained by MQ interactions between FM trimers with imperfect FM directivity
(imperfect FM trimer).
When imperfection of FM directivity exists in the trimer, it can possesses MQ moments.
Because MQ moments possess large degrees of freedom comparing to dipole moments, elimination of frustration easily 
occurs with generation of different kinds of spontaneous MQ moments. When MQ moments with different symmetries
generate on the triangular lattice, geometrical frustration is eliminated because no MQ interactions
work between the moments with different symmetries
\cite{Nakamura2023}.
As a result, the trimers are separated into two sub-groups which do not interact with each other in quadrupole manner.
Thus, the coexistence of different kinds of magnetic anisotropies are generated.
However, the microscopic origin of the MQ interactions in Gd$_{3}$Ru$_{4}$Al$_{12}$ is still an open problem.
Generally speaking, magnetic multipole interactions may be difficult to explain in terms of the RKKY 
mechanism. 
In most cases, the wave numbers of spin polarization correspond to multiple interactions are required to be much higher
than Fermi wave 
number $k_{\rm F}$, although wave numbers of conduction electrons are $\sim k_{\rm F}$ at most. 
In the case of Gd$_{3}$Ru$_{4}$Al$_{12}$, however, the trimers can be regarded as magnetic impurities with long 
diameter effectively. In this case, MQ interactions of the RKKY mechanism may be enhanced.
The multiple spin polarization generated by a trimer may induce multiple spin polarization on another trimer.
These multiple interactions would reflect whole spin structures of the interacting trimers because of their long reaching 
distances.

From another point of view, Gd$_{3}$Ru$_{4}$Al$_{12}$ is one of a host materials of skyrmion with cetrosymmetric
crystal structure. 
In 2019, resonant X-ray diffraction (RXD) measurements were performed on Gd$_{3}$Ru$_{4}$Al$_{12}$
by Matsumura {\it et al.} and Hirschberger {\it et al.}
\cite{Matsumura2019,Hirschberger2019_2}. 
A partial order in the IMT phase II and the formation of FM trimers in the ordered phases 
are supported by these microscopic measurements, and they reported that the
$\bm{S_{r}}$s have a helical structure along the $a$ axis at zero field.
This structure translates into a conical structure of ${\bm S_{r}}$s by applying magnetic fields directed along the $c$ axis,
and SkL is observed in the vicinity of the boundary of these phases
\cite{Hirschberger2019_2}. 
The diameter of the skyrmion observed in Gd$_{3}$Ru$_{4}$Al$_{12}$ is very small, and this would be preferable characteristic 
of high density recordings. 
Conventionally, Dzyaloshinskii–Moriya interaction (DMI) is believed to be an important
element of skyrmion
\cite{Muhlbauer2009,Everschor-Sitte2018},  
however, skyrmions have been observed even in crystals with cetrosymmetric symmetry
where DMI is not expected usually 
\cite{Hirschberger2019_2,Kurumaji2019}. 
Much effort has been made to clarify the mechanisms of stabilizing skyrmions
\cite{Okubo2012,Nagaosa2013,Khanh2020,Wang2020,Yambe2021,Utesov2021,Hayami2021,Paddison2022,Hayami2022}. 
In 2023, it is pointed out that MQ interactions possibly assist to generate the helical spin structure
and the helical-TC phase transition in Gd$_{3}$Ru$_{4}$Al$_{12}$
\cite{Nakamura2023}. 
However, contribution of MQ moments to stabilize SkL is still an open problem.

In the present paper, 
we discuss the MQ moments and chiralities of the trimers in Gd$_{3}$Ru$_{4}$Al$_{12}$.
We deduce MQ interactions of the RKKY mechanism, and propose a Hamiltonian.
Then, we examine consistency between the Hamiltonian and results of experimental studies about
magnetic anisotropies and phase transitions of Gd$_{3}$Ru$_{4}$Al$_{12}$.
Further, we discuss the consistency between the Hamiltonian and spin structures of SkL. 
The Gd$^{3+}$ ion possesses only a pure spin ($S=7/2$), and the large quantum number of this spin
enables to consider it as a classical spin approximately. These features of Gd ions make theoretical treatments simple
combined with strong localized characters of 4$f$ electrons.
Gd$_{3}$Ru$_{4}$Al$_{12}$ is a nice spin system for studying macroscopic magnetic features and microscopic 
spin structure of SkL systematically.

\section{Spin structures in the trimers} 
	\subsection{Symmetric MQ moments in the hexagonal crystal} 

In this section, we discuss symmetric MQ moments and corresponding spin structures of the imperfect FM trimers.
We require these spin structures to have the following additional properties: (1) the resultant spin of the trimers
possess finite value (2) they possess independent degrees of 
magnetic dipole and MQ freedom (3) chiralities can be defined for the structures.
These properties of the trimers make description of the interactions between trimers simple.
In the present paper, we treat dipole moments, spins and MQ moments as classical moments because the quantum 
number of spin on the Gd$^{3+}$ ion is large $(S=7/2)$. However, divalency of spins shall be considered.

\begin{figure}[t]
		\includegraphics[width=8cm]{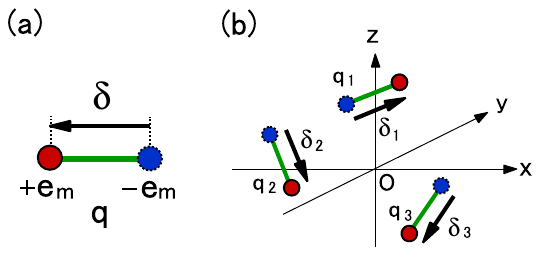}	
		\caption{(a) Magnetic dipole moment ${\bm q}$. The symbols $\pm e_{m}$ and ${\bm \delta}$ denote quantities of 
		magnetic charges and displacement vector from the minus charge to the plus charge, respectively.
		 (b) Magnetic moments ${\bm q_{i}}$ ($i=1,2,3$) distributed in the three dimensional space represented
		  by the coordinates $x$--$y$--$z$. ${\bm \delta_{i}}$ is the displacement vector of each moment.
		}
		\label{f2}
	\end{figure}

Figure~\ref{f2}(a) displays a magnetic dipole moment ${\bm q}$. In this figure, two magnetic charges with quantities $\pm e_{m}$
are placed at the ends of the line segment. The vector ${\bm \delta}$ indicates the displacement vector from
the minus charge to the plus charge. The magnitude of the magnetic dipole moment is described as
		\begin{align}
			{\bm q}=e_{m}{\bm \delta}
			\label{s2_e1a}
		\end{align}
in the unit of ${\rm Wb\,m}$.
This magnetic moment is described as
		\begin{align}
			{\bm \mu}=\mu_{0}^{-1} e_{m}{\bm \delta}
			\label{s2_e2b}
		\end{align}
in the unit of ${\rm A\,m^{2}}$ or of ${\rm JT^{-1}}$, where $\mu_{0}$ is the vacuum permeability.

Symmetric MQ moments in hexagonal crystals are defined by the following formula.
		\begin{align}
			\tilde{Q}_{\Gamma \gamma}= \mu_{0}^{-1} \int \rho_{m} ({\bm r})
				 F_{\Gamma \gamma}({\bm r} ) \, d{\bm r}.  	
			\label{s2_e3c}
		\end{align}
Here, $\rho_{m}$ is the magnetic charge density in the unit of ${\rm Wb\,m^{-3}}$.
The symbol ``tilde'' on the $Q_{\Gamma\gamma}$ in Eq.~\ref{s2_e3c} denotes that the quantity is a dimensional quantity.
The formulas $F_{\Gamma \gamma}({\bm r_{i}})$ in this equation are defined as
		\begin{align}
			F_{2}^{0}({\bm r})&=\frac{1}{2}(2z^{2}-x^{2}-y^{2}),\label{s2_e4d}\\
			F_{2}^{2}({\bm r})&=\frac{1}{2}(x^{2}-y^{2}),	\label{s2_e5e}\\
			F_{xy}({\bm r})&=xy,  \label{s2_e6f}\\
			F_{yz}({\bm r})&=yz,  \label{s2_e7g}\\	
			F_{zx}({\bm r})&=zx,   \label{s2_e8a} 		
		\end{align}
(Appendix A) in the unit of $m^{2}$.  
In the case of the trimer spins, each spin can be replaced by Eq.~\ref{s2_e1a} using tiny displacement vector
 ${\bm \delta}$. Then, Eq.~\ref{s2_e3c} is rewritten as,
		\begin{align}
			\tilde{Q}_{\Gamma \gamma}&= \mu_{0}^{-1} \sum_{i=1}^{3}
			\sum_{\pm} \left[ \pm e_{m} \,
				 F_{\Gamma \gamma} \left({\bm r_{i}} \pm \frac{1}{2} {\bm \delta_{i}} \right)  \right] \label{s2_e9b} \\
			&= \mu_{0}^{-1} \sum_{i=1}^{3} e_{m} \, {\bm \delta_{i}}  \cdot
			\biggl[ \frac{\partial F_{\Gamma \gamma}({\bm r})}  {\partial {\bm r} }   \biggr]_{\bm r= \bm r_{i}} \label{s2_e10c} \\
				&= \sum_{i=1}^{3} {\bm \mu_{i}}  \cdot
			\left[ \frac{\partial F_{\Gamma \gamma}({\bm r})}  {\partial {\bm r} } \right]_{\bm r= \bm r_{i}}  \label{s2_e11d}\\
				&=-g_{s}  \mu_{\rm B}\sum_{i=1}^{3} {\bm S_{i}}  \cdot 
					\left[{\bm \nabla} F_{\Gamma \gamma}({\bm r}) \right]_{\bm r= \bm r_{i}}.
				\label{s2_e12e}
		\end{align}
Here, the symbol ${\bm \mu_{i}}$ is the magnetic moment of Gd$^{3+}$ ions, $\mu_{\rm B}$ Bohr magneton, and $g_{s}$ the 
$g$-factor, respectively. 
In Eqs.~\ref{s2_e9b}-\ref{s2_e12e}, the double sign corresponds, and
$\tilde{Q}_{\Gamma \gamma}$ is described in the unit of ${\rm JT^{-1}m}$ or unit of $\mu_{\rm B}{\rm m}$.
The formula $F_{\Gamma \gamma}({\bm r} )$ look like potential functions outwardly in these equations. 
In Eq.~\ref{s2_e12e}, each ${\bm S_{i}}$ is directed along ${\bm \nabla} F_{\Gamma \gamma}({\bm r_{i}})$.
In Eq.~\ref{f12}, the directions of spins are fixed in the real space, therefore, the degrees of freedom due to
divalency of spins would translate to the degrees of freedom of $\tilde{Q}_{\Gamma\gamma}$.

	\begin{figure}[b]
		\includegraphics[width=8cm]{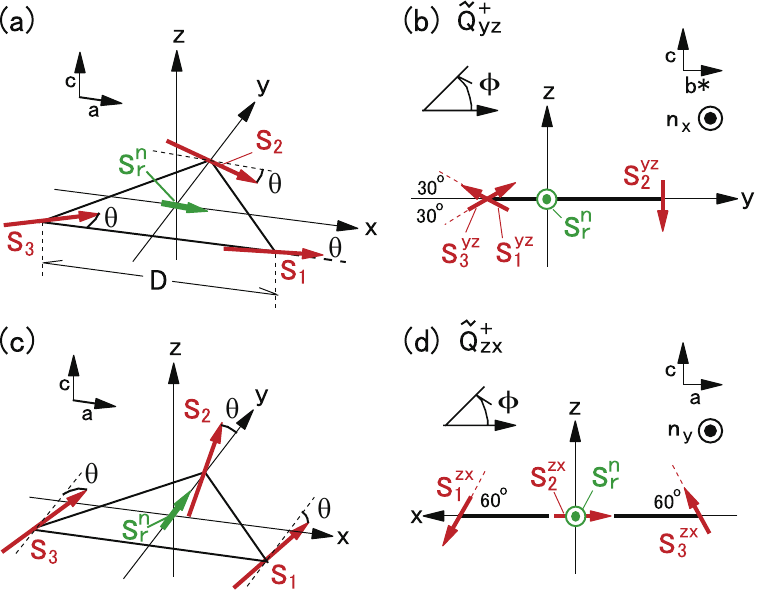}	
		\caption{Spin structures of the imperfect FM trimers.
		The coordinates of $x$, $y$ and $z$ are taken parallel to the $a$, $b^{*}$ and $c$ axes,
		respectively.
		${\bm S_{i}}$ $(i=1,2,3)$ are spins of Gd ions. The unit vectors ${\bm S_{r}^{n}}$s denote the 
		directions of the resultant spins ${\bm S_{r}}$s. ${\bm S_{i}^{yz}}$ and ${\bm S_{i}^{zx}}$ denote the $yz$ plane
		components and $zx$ plane components of  ${\bm S_{i}}$s ($i=1,2,3$), respectively. 
		The symbol $\phi$ represents the angle around the quantized axis (see text).
		The unit vectors ${\bm n_{l}}$ ($l=x,y,z$) represent those in Eq.~\ref{s2_e19b}.
		(a) The symbol $\theta$ denotes the angle between the $x$ axis and each component spin, respectively. 
		(b) Project drawing of (a) with respect to the $x$ axis. 
		The sign $+$ on the right shoulder of $\tilde{Q}_{yz}$ denotes the sign of the chirality.
		(c) Trimer of which ${\bm S_{r}}$ is directed along the $y$ axis. The symbol $\theta$ denotes the angle between 
		the $y$ axis and each component spin, respectively. 
		(d) Project drawing of (c) with respect to the $y$ axis. 
		 The symbol ``$+$'' on the right shoulder of $\tilde{Q}_{yz}$ denotes the sign of the chirality.
		}
		\label{f3}
	\end{figure}
	\begin{figure}[h]
		\includegraphics[width=8cm]{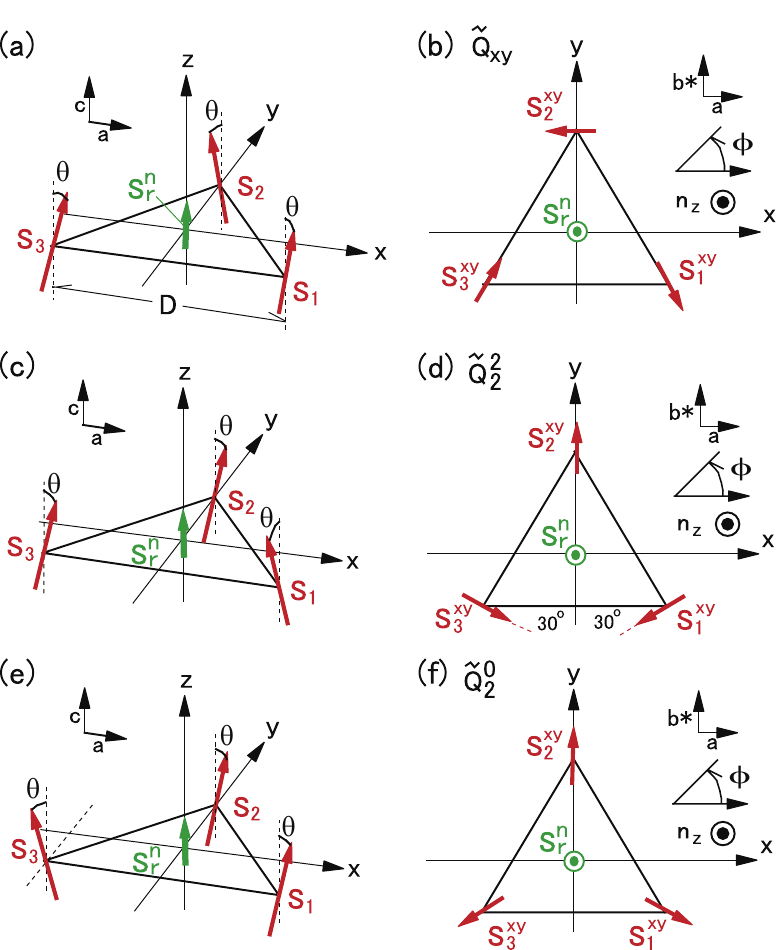}	
		\caption{Spin structures of the imperfect FM trimers.
		The coordinates of $x$, $y$ and $z$ are taken parallel to the $a$, $b^{*}$ and $c$ axes,
		respectively. ${\bm S_{i}^{xy}}$ denote the $xy$ plane components of  ${\bm S_{i}}$ ($i=1,2,3$).
		${\bm S_{i}}$ $(i=1,2,3)$ are spins of Gd ions. The unit vectors ${\bm S_{r}^{n}}$s denote the 
		directions of the resultant spins ${\bm S_{r}}$s. 
		$\phi$ is the angle around the quantized axis (see text).
		The unit vectors ${\bm n_{l}}$ ($l=x,y,z$) represent those in Eq.~\ref{s2_e19b}.
		(a) Trimer which possesses $\tilde{Q}_{xy}$. $\theta$ denotes the angle between 
		the $z$ axis and each component spin, respectively. 
		(b) Project drawing of (a) with respect to the $z$ axis.
		(c) Trimer which possesses $\tilde{Q}_{2}^{2}$.  $\theta$ denotes the angle between 
		the $z$ axis and each component spin, respectively. 
		(d) Project drawing of (c) with respect to the $z$ axis. 
		(e) Trimer which possesses $\tilde{Q}_{2}^{0}$.  $\theta$ denotes the angle between 
		the $z$ axis and each component spin, respectively. 
		(f) Project drawing of (e) with respect to the $z$ axis. 
		}
		\label{f4}
	\end{figure}
%

	\subsection{Spin structures of the trimers which possess symmetric MQ moments} 
The resultant spin at the trimer is defined as
		\begin{align}
			\bm{S_{r}}={\bm S_{1}}+{\bm S_{2}}+{\bm S_{3}}.
			\label{s2_e13f}
		\end{align}
Here, ${\bm S_{i}}$ ($i=1,2,3$) is the component spin on each Gd$^{3+}$ ion which belongs to the trimer.
When temperature is high, the spin structure of the trimer is described using
${\bm S_{r}}$ $(S_{r}=21/2)$. 
However, FM trimers are deformed by the MQ interactions at low temperatures.
The model of the spin structure ($T<T_{1}$) determined by RXD 
insists the imperfect FM directivity on each trimer
\cite{Matsumura2019}.
This implies that another spin structures of the trimers is realized at low temperatures. We assume this set of symmetric 
spin structures as illustrated in Fig.~\ref{f3} and Fig.~\ref{f4} with reference to the previous study
\cite{Nakamura2023}.
These spin structures possess independent degrees of freedom of magnetic dipole and MQ moments.
In Figs.~\ref{f3} and \ref{f4}, the coordinate axes $x$, $y$ and $z$ are directed along the $a$, $b^{*}$ and $c$ axes, respectively.
The origin of the $x$--$y$--$z$ coordinate is taken at the center of the gravity of the triangle, and
the $ab$ plane corresponds to the plane of $z=0$. The symbol $D$ in Figs.~\ref{f3} and \ref{f4} is the length of the side of the
smaller triangle illustrated in Fig.~\ref{f1}(c). 
The red arrows ${\bm S_{i}}$ $(i=1,2,3)$ in Fig.~\ref{f3}(a), \ref{f3}(c) and Fig.~\ref{f4}(a), \ref{f4}(c), \ref{f4}(e) denote 
the spins on Gd$^{3+}$ ions ($i$ is the vertex number), and
the green arrows ${\bm S_{r}^{n}}$s indicate unit vectors parallel to the ${\bm S_{r}}$s.
We consider only the smallest angle $\theta$ which is allowed by quantum nature (Appendix B). 
This assumption is reasonable because both $T_{1}$ and $T_{2}$ are much smaller than the binding energy of FM trimer.
Figures~\ref{f3}(b) and \ref{f3}(d) are project drawings of Figs.~\ref{f3}(a) and \ref{f3}(c) with respect to the $x$ axis and 
$y$ axis, respectively.
In Figs.~\ref{f4}(b), \ref{f4}(d) and \ref{f4}(f) which are project drawings of  Fig.~\ref{f4}(a), \ref{f4}(c), \ref{f4}(e) 
with respect to the $z$ axis are respectively presented.
The red arrows in ${\bm S_{i}}^{\alpha}$ $(\alpha=yz, zx, xy)$ in Figs.~\ref{f3} and \ref{f4}
are $\alpha$ plane components of  ${\bm S_{i}}$.

	\begin{figure}[b]
		\includegraphics[width=8cm]{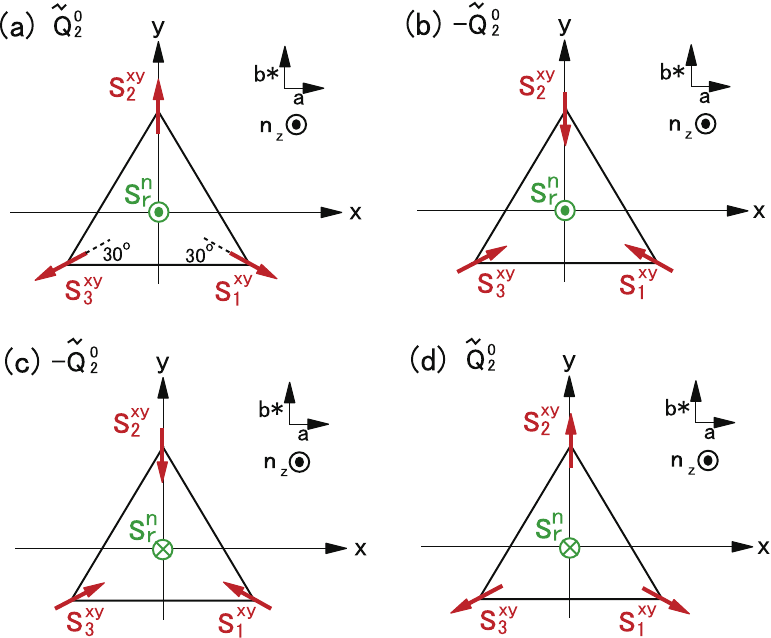}	
		\caption{(a) Spin structure of the trimer which possesses $\tilde{Q}_{2}^{0}$. 
		(b) The spin structure in which ${\bm S_{i}}^{xy}$s in panel (a) turn to the opposite directions.
		The sign of MQ moment changes to the opposite sign.
		(c) The spin structure in which the spin structure in panel (a) is time reversed.
		(d) The spin structure in which only the ${\bm S_{r}}$ in panel (a) turns to the opposite direction.
		The unit vectors ${\bm n_{l}}$ ($l=x,y,z$) represent those in Eq.~\ref{s2_e19b}.
		The sign of MQ moment does not change.
		}
		\label{f5}
	\end{figure}

The ${\bm S_{i}^{\alpha}({\bm r_{i}})}$ carries the magnetic moment
		\begin{align}
			{\bm \mu_{i}^{\alpha}}=-g_{s}\mu_{\rm B} \bm S_{i}^{\alpha}({\bm r_{i}},\phi_{i}),	
			\label{s2_e14g}	
		\end{align}
where $g_{s}=2$ is the $g$-value of the spin. Directions of ${\bm S_{i}}^{\alpha}$ are perpendicular to ${\bm S_{r}}$
		\begin{align}
			{\bm S_{i}}^{\alpha}({\bm r_{i}},\phi_{i}) \perp {\bm S_{r}}.	
			\label{s2_e15h}	
		\end{align}
Here, $\phi_{i}$ is the angle in the $\alpha$ plane defined in Figs.~\ref{f3} and \ref{f4}. 
The reference directions of $\phi_{i}=0$ are
		\begin{align}
			&{\bm n_{x}} &({\bm S_{r}^{n}} \parallel z), 		\label{s2_e16i} \\
			&{\bm n_{z}}\times {\bm S_{r}^{n}}  &({\bm S_{r}^{n}} \perp z), \label{s2_e17j}				
		\end{align}
as shown in Figs.~\ref{f3} and \ref{f4}. The angle $\phi_{i}$ has the meaning in that it is the angle around the quantized axis.

Spin structures 
in Figs.~\ref{f3} and \ref{f4} possess MQ moments.
We arrange these MQ moments $\tilde{Q}_{\Gamma \gamma}$s in Tabel~\ref{Tab1}. 
When all of the 
${\bm S_{i}^{\alpha}} ({\bm r_{i}})$ in each spin trimer turn into the opposite directions simultaneously, the sign of MQ moment
changes into the opposite sign. In Fig.~\ref{f5}(a) and \ref{f5}(b), two kinds of spin trimers are
illustrated.
The spin trimer in Fig.~\ref{f5}(a) possesses MQ moment $\tilde{Q}_{2}^{0}$. 
In Fig.~\ref{f5}(b),
the directions of ${\bm S_{i}^{xy}}$s are in the opposite directions comparing to those in Fig.~\ref{f5}(a), respectively.
Then, the trimer in this figure possesses $-\tilde{Q}_{2}^{0}$.

	\begin{figure}[b]
		\includegraphics[width=8cm]{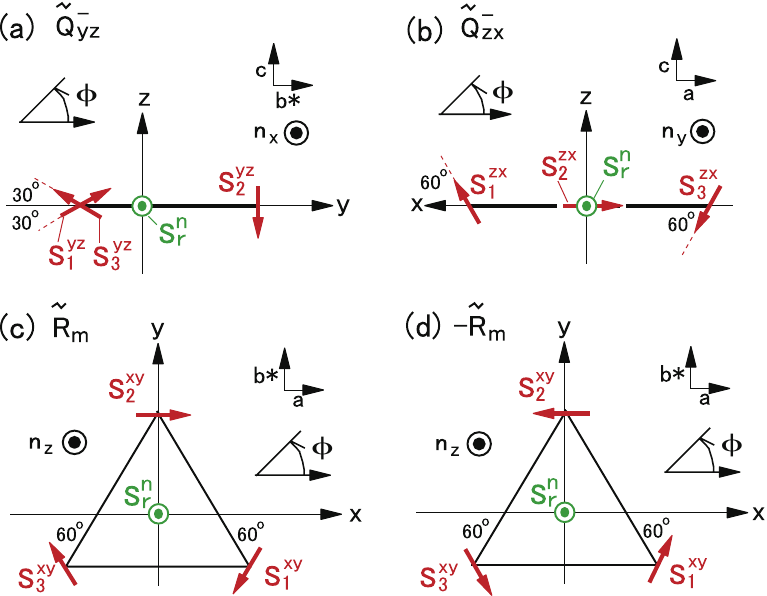}	
		\caption{(a) Spin structure of $\tilde{Q}_{yz}^{-}$ which has the opposite sign of chirality of $\tilde{Q}_{yz}^{+}$ in 
		Fig.~\ref{f3}(b).
		The sign $\phi$ denotes the angle around ${\bm S_{r}^{n}}$ from the reference direction indicated by the arrow
		(see text). The unit vectors ${\bm n_{k}}$ ($k=x,y,z$) represent the directions of quantized axes of ${\bm S_{i}}$.
		(b) Spin structure of $\tilde{Q}_{zx}^{-}$ which has the opposite sign of chirality of $\tilde{Q}_{zx}^{+}$ in Fig.~\ref{f3}(d).
		(c) Rotational spin structure of the trimer where  ${\bm S_{i}^{xy}}$s turn to the clockwise
		with increasing $i$. 
		(d) Rotational spin structure of the trimer where  ${\bm S_{i}^{xy}}$s turn to the counterclockwise
		with increasing $i$.
		The unit vectors ${\bm n_{l}}$ ($l=x,y,z$) represent those in Eq.~\ref{s2_e19b}.
		}
		\label{f6}
	\end{figure}

	\renewcommand{\arraystretch}{1.5}			
					
		\begin{table*}[t]
			\caption{Classical MQ moments $\tilde{Q}_{\Gamma \gamma}$, rotational moment $\tilde{R}_{m}$, scalar 
			chiralities $C_{sc}$ 
			carried by the trimers depicted in Figs.~\ref{f3}, \ref{f4}, \ref{f5} and \ref{f6} in the classical spin approximation
			$|S_{i}|=7/2$ (Appendix B).
			The symbol $\xi$ is defined as $\sin{\theta}=0.6999$ (Eq.~\ref{s2_e22e}). This table includes functions 
			$F({\bm r})s$ defined in Eqs.~\ref{s2_e4d}-\ref{s2_e8a} and Eq.~\ref{s2_e21d}. The length of the side of the small
			triangle $D$ is 0.3706 nm.
			}							
			\center					
			\begin{tabular}{ccccccccccc}
			\hline 
			spin trimer & $\tilde{Q}_{yz}^{+}$  & $\tilde{Q}_{yz}^{-}$ & $\tilde{Q}_{zx}^{+}$ & $\tilde{Q}_{zx}^{-}$  
			& $\tilde{Q}_{xy}$& $\tilde{Q}_{2}^{2}$ & $\tilde{Q}_{2}^{0}$ &$\tilde{R}_{m}$ & $C_{sc}$ & $F({\bm r})$\\
			\hline
			in Fig.~\ref{f3}(b)\, & $\frac{7\sqrt{3}}{2} D \xi \mu_{\rm B} $ 
			& 0 & 0 & 0 & 0 &0 &0 &0 & $+\xi^{2}$ & $F_{yz}$ \\
			in Fig.~\ref{f3}(d)\, 
			& 0 & 0 &$\frac{7\sqrt{3}}{2} D \xi \mu_{\rm B} $  & 0 & 0 &0 &0 &0
			 & $+\xi^{2}$ & $F_{zx}$ \\
			in Fig.~\ref{f4}(b)\, & 0 
			& 0 & 0 & 0 &  $7\sqrt{3} D \xi \mu_{\rm B} $ &0 &0 &0 & $-\xi^{2}$ & $F_{xy}$  \\
			in Fig.~\ref{f4}(d)\, & 0 &  0 
			& 0 & 0 & 0 & $7\sqrt{3} D \xi \mu_{\rm B} $ &0 &0 & $-\xi^{2}$ & $F_{2}^{2}$  \\
			in Fig.~\ref{f4}(f)\, & 0 & 0 & 0
			& 0 & 0 & 0  & $7\sqrt{3} D \xi \mu_{\rm B} $ &0&  $+\xi^{2}$ & $F_{2}^{0}$  \\
			in Fig.~\ref{f5}(b)\, & 0 & 0 & 0 &0 & 0 &0  
			&$-7\sqrt{3} D \xi \mu_{\rm B} $ &0 & $+\xi^{2}$ & $F_{2}^{0}$  \\
			in Fig.~\ref{f6}(a)\, & 0 &  $\frac{7\sqrt{3}}{2} D \xi\mu_{\rm B}$ & 0 &0 & 0 &0 
			&0 &0 & $-\xi^{2}$ & $F_{yz}$  \\
			in Fig.~\ref{f6}(b)\, & 0 & 0 & 0 &  $\frac{7\sqrt{3}}{2} D \xi \mu_{\rm B}$  & 0 &0 &0 &0
			&$-\xi^{2}$ & $F_{zx}$  \\
			in Fig.~\ref{f6}(c)\, & 0 & 0 & 0 & 0  & 0 &0 &0 &$7\sqrt{3} D \xi \mu_{\rm B}$ &$+\xi^{2}$& ${\bm F_{R}}$\\
			in Fig.~\ref{f6}(d)\, & 0 & 0 & 0 & 0  & 0 &0 &0 &$-7\sqrt{3} D \xi \mu_{\rm B}$ &$+\xi^{2}$& ${\bm F_{R}}$\\
			\hline 
			\end{tabular}
			\label{Tab1}
		\end{table*}											
															
	\renewcommand{\arraystretch}{1}

	\begin{figure}[b]
		\includegraphics[width=8cm]{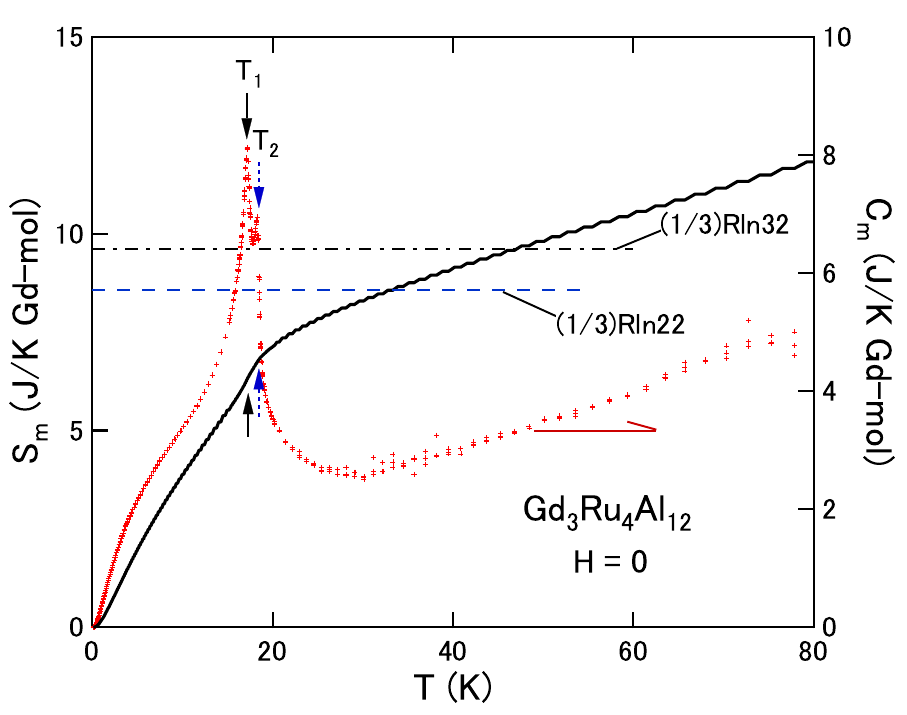}	
		\caption{Magnetic entropy (the solid line) and magnetic specific heat (the red crosses)
		of Gd$_{3}$Ru$_{4}$Al$_{12}$ at zero field. Data is taken from the previous paper
\cite{Nakamura2023}.
		The pair of solid black arrows and the pair of dotted blue arrows denotes $T_{1}$ and $T_{2}$, respectively.
		The broken blue line and dotted-broken black line are indications of the values 
		$S_{m}=R\ln{22}$ and $R\ln{32}$,
		respectively. Here, $R$ is the gas constant.
		}
		\label{f7}
	\end{figure}

	\subsection{Chiralities and rotational magnetic moments} 
When spins are placed on the vertex of a triangle, they possess chiralities in general.
Chiralities in magnetic systems are usually defined as the vector spin chiralities
\cite{MiyashitaShiba1984}
		\begin{align}
			{\bm C_{vs}}=\frac{2}{3\sqrt{3}} \sum_{(i,j)} ({\bm S_{i}\times {\bm S_{j}}}),
			\label{s2_e18a}
		\end{align}
where  $(i,j)$ runs over $(1,2),(2,3),(3,1)$. According to this definition, the chiralities are given as vectors.
In this section,
we define the scalar chiralities of the trimer as  
		\begin{align}
			C_{sc}=\frac{8}{147\sqrt{3}}\sum_{(i,j)} ({\bm S_{i}^{\alpha}\times {\bm S_{j}}^{\alpha}}) \centerdot {\bm n_{l}}
			\label{s2_e19b}
		\end{align}
for the sake of convenience of discussion
\cite{Barron2012}.
Here, $(i,j)$ runs over $(1,2),(2,3),(3,1)$, and ${\bm n_{l}}$ ($l=x, y, z)$ are the unit vectors directed along the 
fundamental unit vectors
of x--y--z coordinate. Figures~\ref{f3}-\ref{f5} include the indications of the directions of  ${\bm n_{l}}$s.
If we take chiralities into consideration, there are other degrees of freedom the trimer.
In Fig.~\ref{f5}(c), the spin structure is time reversed to that in Fig.~\ref{f5}(a), then the sign of 
$\tilde{Q}_{2}^{0}$ in Fig.~\ref{f5}(c) is opposite to that in Fig.~\ref{f5}(a),
but the signs of $C_{sc}$ in Fig.~\ref{f5}(a) and \ref{f5}(c) are the same. 
In Fig.~\ref{f5}(d), only ${\bm S_{r}}$ ($ S_{r}=15/2$) is changed into the opposite direction of ${\bm S_{r}}$ in Fig.~\ref{f5}(a).
In this case, neither signs of $\tilde{Q}_{2}^{0}$ nor of $C_{sc}$ are not changed.
Thus, the spin trimer which possesses $\tilde{Q}_{2}^{0}$ has 4 degrees of freedom. 
Each trimer which possesses MQ moment also has 4 degrees of freedom. 
Figure~\ref{f6}(a) and \ref{f6}(b)
show spin structures which possess $\tilde{Q}_{yz}$ and $\tilde{Q}_{zx}$, respectively, 
Both the signs of chiralities of these structures are minus.
On the other hand, both the signs of chiralities of spin structures in Fig.~\ref{f3}(b) and \ref{f3}(d) are plus. 
The signs ``$\pm$'' 
which are placed on upper right side of $\tilde{Q}_{yz}$ and $\tilde{Q}_{xz}$ in Figs.~\ref{f3} and \ref{f6} 
denote the signs of $C_{sc}$.

Figure~\ref{f6}(c) denotes  a rotational spin structure
of ${\bm S_{i}^{xy}({\bm r_{i}})}$s in the $xy$ plane. 
We define rotational magnetic (RM) moment as
		\begin{align} 
			\tilde{R}_{m}&=-g_{s}\mu_{\rm B}\sum_{i=1}^{3} {\bm S_{i}}^{xy} \cdot  \left[ {\bm \nabla} \times  
			{\bm F_{R}} \left({\bm r} \right) \right]_{{\bm r}={\bm r_{i}}}.
					\label{s2_e20c}
		\end{align}
Here, ${\bm F_{R}}$ is a vector potential like function defined as
		\begin{align} 
			{\bm F_{R}} ({\bm r})&=
				\begin{pmatrix}
					xz					&			yz		&		0	
				\end{pmatrix}.
					\label{s2_e21d}
		\end{align}
The symbol ``tilde'' on $R_{m}$ indicates that $\tilde{R}_{m}$ is dimensional quantity.
The spin structures in Fig.~\ref{f6}(c) has no MQ moments. 
When all ${\bm S_{i}^{xy}({\bm r_{i}})}$s turn to opposite directions at once, 
the rotational structure in Fig.~\ref{f6}(c) replace to the structure in Fig.~\ref{f6}(d) and the sign of 
$\tilde{R}_{m}$ changes to negative.
Each rotational state possesses two degrees of freedom concerning the directions of ${\bm S_{r}}$.
Therefore, the trimer which possesses a $\tilde{R}_{m}$ have 4 degrees of freedom as a whole. 
Consequently, the spin structures of trimers with MQ moments or RM moments have 32 degrees of freedom as a whole.
We summarized $\tilde{Q}_{\Gamma \gamma}$, $\tilde{R}_{m}$, $C_{sc}$ of spin trimers in Table~\ref{Tab1}.
As shown in Table~\ref{Tab1}, both the MQ moments and RM moments are proportional to $D$, and the interactions
between these moments are proportional to $D^{2}$.
The length of $D$ is much greater than the diameter of 4$f$ electron in Gd$^{3+}$. 
This is a major reason why MQ moments and MQ interactions cannot be ignored in Gd$_{3}$Ru$_{4}$Al$_{12}$
with the breathing kagome structure 
\cite{Nakamura2023}.
According to previous study, the lattice constants are 0.8778 nm for $a$ axis, 0.9472 nm for the $c$ axis, and
$D$ is obtained to be 0.3706 nm.
The symbol $\xi$ in Table~\ref{Tab1} is defined as
		\begin{align} 
			\xi= \sin{\theta}=0.6999
					\label{s2_e22e}
		\end{align}
(Appendix B).

\begin{figure}[b]
		\includegraphics[width=8cm]{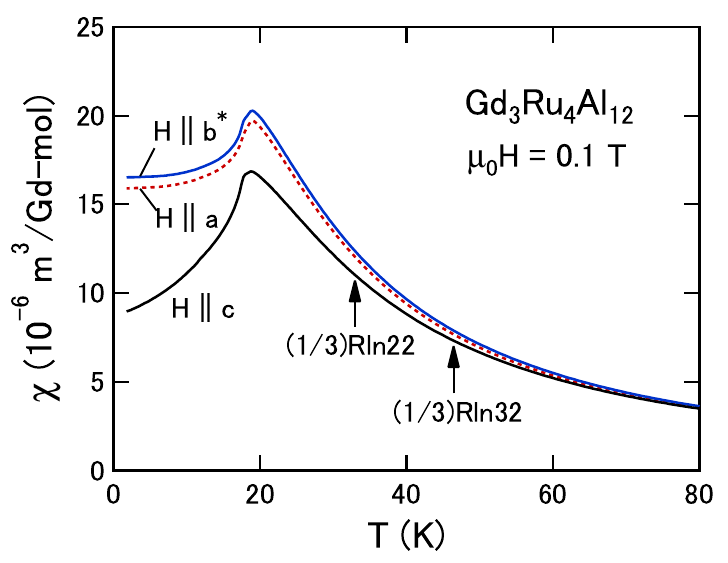}	
		\caption{Magnetic susceptibility of Gd$_{3}$Ru$_{4}$Al$_{12}$ at 0.1 T. Data is taken from the
		 previous paper
\cite{Nakamura2023}.
		The dotted red, solid blue, and solid black lines denote magnetic susceptibilities
		for $H\parallel a$, $H\parallel b^{*}$, and $H\parallel c$, respectively.
		The downward arrow corresponds to the temperatures $T_{s}$. The upward arrows correspond
		to the temperatures at which $S_{m}$ is equal to $R\ln{22}$ and $R\ln{32}$, respectively (see Fig.~\ref{f7}). 
		}
		\label{f8}
	\end{figure}

We present the magnetic entropy $S_{m}$ and magnetic specific heat $C_{m}$ of Gd$_{3}$Ru$_{4}$Al$_{12}$ at zero field 
in Fig.~\ref{f7}. Data in this figure is taken from the previous paper
\cite{Nakamura2023}.
The broken blue line and the dotted-broken black line in Fig.~\ref{f7} are the indications of the values $S_{m}=(1/3)R\ln{22}$ and
 $S_{m}=(1/3)R\ln{32}$, respectively.
The value $(1/3)R\ln{22}$ is the lower limit of $S_{m}$ which is expected from paramagnetic state
of perfect FM trimers ($S_{r}=21/2$).
As shown in Fig.~\ref{f7}, $C_{m}$ begins to increase rapidly with decreasing temperature when $S_{m}$ reaches
to this value.
The value $(1/3)R\ln{32}$ is the upper limit of $S_{m}$ which is expected from the imperfect FM trimers
mentioned above.
Figure~\ref{f8} presents the anisotropy in the magnetic susceptibilities $\chi$ of  Gd$_{3}$Ru$_{4}$Al$_{12}$. 
Data in this figure is taken from the previous paper
\cite{Nakamura2023}.
As shown in Fig.~\ref{f8}, slight anisotropy begins to appear approximately below 50 K. 
Because the magnetic anisotropy is not expected from the perfect FM trimerization,
the magnetic anisotropy below 50 K in paramagnetic phase may be related to the generation of imperfect FM trimers.
This temperature approximately corresponds to the $S_{m}=R{\rm ln}32$ ($T=47$ K) as shown in Fig.~\ref{f8}.
The specific heat and magnetic susceptibility implies the generation of the component of MQ moments of the spin trimers
occurs in the temperature range $35\lesssim T \lesssim 50$ K with decreasing temperature.

%

\section{MQ interactions and rotational spin interactions of the RKKY mechanism}

	\subsection{Dimensionless Quadrupole spins, rotational spins and spin structures of the trimers} 

We discussed $\tilde{Q}_{\Gamma\gamma}$s and $\tilde{R}_{m}$ moments in the previous section, however, if no interactions 
work between them, these moments are not realized. The multiple moments
and the mechanism of multiple interactions would be inseparable research subjects. 
In this section, we define dimensionless description of the trimer states obtained in the previous section and discuss
the origin of the MQ interactions according to the RKKY mechanism. After that, we propose a
Hamiltonian which is described by the multiple interactions between the trimers.

For the sake of improving outlook of calculation, we define dimensionless quadrupole spin (QS) as
		\begin{align}
			 Q_{\Gamma \gamma}&= D^{-1} \sum_{i=1}^{3} 
				 {\bm S_{i}^{\alpha}} (\Gamma\gamma,\phi_{i}) \cdot \left[ \nabla 
				 F_{\Gamma \gamma}({\bm r}) \right]_{\bm r= \bm r_{i}}.
				\label{s3_e23ff}
		\end{align}
Here, 
		\begin{align}
			 {\bm S_{i}^{\alpha}}(\Gamma\gamma,\phi_{i}) \parallel 
			 \left[\nabla F_{\Gamma \gamma}({\bm r})\right]_{\bm r= \bm r_{i}},
				\label{s3_e24ffGGG}
		\end{align}
and the directions of these vectors are the same.
Similarly, we define dimensionless rotational spin (RS) as
		\begin{align}
			R_{m}= D^{-1} \sum_{i=1}^{3}  {\bm S_{i}^{xy}} (R_{m},\phi_{i})   \cdot \left[ {\bm \nabla} \times 
			{\bm F_{R}} \left({\bm r} \right) \right]_{\bm r= \bm r_{i}},
				\label{s3_e25gg}
		\end{align}
where
		\begin{align}
			 {\bm S_{i}^{xy}}(R_{m},\phi_{i}) \parallel 
			 \left[{\bm \nabla} \times {\bm F_{R}({\bm r}})\right]_{\bm r= \bm r_{i}},
				\label{s3_e26hHHHh}
		\end{align}
and the directions of these vectors are the same.

Referring to spin structures of trimers mentioned before, we summarize spin structures in Table~\ref{Tab2}
making them into expressions as follows.
		\begin{align}
			{\bm S_{i}}
			=({\bm r}_{i};S_{i}^{x},S_{i}^{y},S_{i}^{z}).
				\label{s2_e27hh}
		\end{align}
Here, ${\bm r_{i}}$ denotes the position where ${\bm S}_{i}$ placed. In table~\ref{Tab2}, the coefficients 
$(7/2)\sin{\theta}=(7/2)\xi$ (Appendix B) in the parentheses are omitted for easy look. 
Because of the strong localization of 4$f$ electrons, overlapping between the wave functions of 4$f$ electrons 
on different ${\bm r_{i}}$ ($i=1,2,3$) can be ignored.
The unit vectors ${\bm n_{l}}$ ($l=x,y,z$) in Figs.~\ref{f3}, \ref{f4}, \ref{f5} and \ref{f6} and are included in Table~\ref{Tab2}.
In this table, the angles $\phi_{i}$ in trigonometric functions are defined as angles shown in 
Figs.~\ref{f3}, \ref{f4}, \ref{f5} and \ref{f6}. 
These angles are described as
		\begin{align}
			\phi_{i}=\frac{n_{i} \pi}{6}\quad(i=1,2,3),
				\label{s2_e28ii}
		\end{align}
where $n_{i}$ are integers. We summarize $n_{i}$ in Table~\ref{Tab2}. 
We put the opposite sign of $\phi_{i}$ when $Q_{\Gamma\gamma}$ or $R_{m}$ shows the opposite sign,
taking divalency of spins into consideration.
Then, the angles $\phi_{i}$ distribute in the range $-2\pi < \phi_{i}  \le2\pi$ as presented in Table~\ref{Tab2}.

In the case of imperfect trimers in Gd$_{3}$Ru$_{4}$Al$_{12}$, definition of chirality using permutations is convenient.
We define the following integer $n_{0}$ and permutation $\sigma$ as
		\begin{align}
			n_{0}&\equiv {\rm min} \{n_{i}\},
				\label{s2_e29cI}			\\
			\sigma&=	   
				\begin{pmatrix}
					0					&			1		&		2				\\	
					 |n_{1}-n_{0}|/4			&	|n_{2}-n_{0}|/4		&	|n_{3}-n_{0}|/4		
				\end{pmatrix}
				,
				\label{s2_e30k}
		\end{align}
respectively.
Here, integers in the upper column $\{0,1,2\}$ in Eq.~\ref{s2_e30k} correspond to the numbers which
subtract 1 from vertex numbers of the triangle $i$.
Then, the sign of the permutation sgn$(\sigma)$ indicates the sign of chirality.
For example, in the case of $Q_{yz}^{+}$, $\sigma$ is given as
		\begin{align}
			 \sigma=
				\begin{pmatrix}
					0					&			1		&		2				\\	
					1					&			2		&		0	
				\end{pmatrix}
			=(0\quad2)(0\quad1),
				\label{s2_e31eE}
		\end{align}
according to Table~\ref{Tab2}. Thus, sgn$(\sigma)=+1$. For another example, in the case of $-Q_{yz}^{-}$,
		\begin{align}
			 \sigma=
				\begin{pmatrix}
					0					&			1		&		2				\\	
					1					&			0		&		2	
				\end{pmatrix}
			=(0\quad1),
				\label{s2_e32Ff}
		\end{align}
and sgn$(\sigma)=-1$ is obtained. We summarize sgn$(\sigma)$ in Table II. The chiralities shown in
Table~\ref{Tab2} are consistent with those shown in Table~\ref{Tab1}.

As shown in Table II, independent variables which describe the states of
the trimers are ${\rm sgn}(m_{l})$, sgn$(\sigma)$ and $n_{2}$. 
Here, $m_{l}$ is the magnetic quantum number of the component spin ${\bm S_{i}}$.
First, ${\rm sgn}(m_{l})$ corresponds to the direction of ${\bm S_{r}}$ whose degrees of freedom is two.
Second, sgn$(\sigma)$ corresponds to the chirality whose degrees of freedom is two.
Third, $n_{2}$ which corresponds to directions of the spins on the second vertex of the triangle in the $\alpha$ plane. 
This angle represents the angle around the quantized axis of the trimer.
Single spin of Gd$^{3+}$ ($S=7/2$, $m=5/2$) has two degrees of angular freedom around the  quantization axis arising from 
divalency of spins. 
Therefore, single trimer possesses $2^{3}=8$ degrees of angular freedom around the quantized axis. 
As we can see in Table~\ref{Tab2}, the states are categorized into 8 types by $n_{2}$ actually. 
Two states categorized to the same $n_{2}$ in this table form a pair of plus and minus chiralities.
It should be noted that $n_{2}$ is not the unique parameter for classifying the trimer states.
Similar classification is possible using mean values of $n_{i}$, for example.
In addition to the states of trimers presented in Table~\ref{Tab2}, there are another set of 16 states corresponding to
 $m_{l}=-5/2$.
Consequently, each trimer possesses 32 degrees of freedom. This is consistent to the discussion about $S_{m}$
in the previous section.

	\renewcommand{\arraystretch}{2.0}

		\begin{table*}[t]
			\caption{Component spins of the trimers for $m_{l}=+5/2$ ($l=x,y,z$) which possess QS or RS. 
			When the spin structures corresponding to the states
			are illustrated in the present paper, figure numbers (FN) are given. The unit vectors ${\bm n_{l}}$ ($l=x,y,z$) 
			in these figures denote the direction of ${\bm S_{r}}$s.
			This table includes angles
			$n_{i}$ (in Eq.~\ref{s2_e28ii}) and ${\rm sgn} (\sigma)$ (in Eq.~\ref{s2_e30k}). 
			Magnitude (mag.)
			of the moments of $ Q_{\Gamma\gamma}$ or $ R_{m}$ are also presented.  
			The coefficients $(7/2)\sin{\theta}=(7/2)\xi$ (Appendix B)
			in the parentheses are omitted for easy look. 
			In this table, we determine the range of angles in trigonometric functions as $-2\pi < \phi \le 2\pi$ 
			considering bivalence of the spins.
			}							
			\center					
			\begin{tabular}{lcclrrrcc}
			\hline
			 QS,RS & FN  &${\bm n_{l}}$& \quad component spins of trimers ($m_{l}=+5/2$) 
			&$n_{1}$ &$n_{2}$ &$n_{3}$ & sgn$(\sigma)$ & mag. \\
			\hline
			$ Q_{yz}^{+}$ &\ref{f3}(b) & ${\bm n_{x}}$ 
			&\, $( {\bm r_{1}};\frac{5}{2}, \cos{\frac{5\pi}{6}}, \sin{\frac{5\pi}{6}})
	 				,( {\bm r_{2}};\frac{5}{2}, \cos{\frac{9\pi}{6}}, \sin{\frac{9\pi}{6}} )
	 				,( {\bm r_{3}};\frac{5}{2}, \cos{\frac{\pi}{6}}, \sin{\frac{\pi}{6}} )$  
	 				& $5$ & $9$ & $1$ & $+1$ & $-\frac{7\sqrt{3}}{4}\xi $ \\
			$ Q_{yz}^{-}$ &\ref{f6}(a) & ${\bm n_{x}}$ 
			&\, $( {\bm r_{1}};\frac{5}{2}, \cos{\frac{\pi}{6}}, \sin{\frac{\pi}{6}})
	 				,( {\bm r_{2}};\frac{5}{2}, \cos{\frac{9\pi}{6}}, \sin{\frac{9\pi}{6}})
	 				,( {\bm r_{3}};\frac{5}{2}, \cos{\frac{5\pi}{6}}, \sin{\frac{5\pi}{6}} )$ 
	 				& $1$ & $9$ & $5$ & $-1$ & $-\frac{7\sqrt{3}}{4}\xi $ \\
	 		$ Q_{zx}^{+}$ &\ref{f3}(d) & ${\bm n_{y}}$ 
	 		&\, $( {\bm r_{1}}; \cos{\frac{8\pi}{6}}, \frac{5}{2}, \sin{\frac{8\pi}{6}} )
	 				,( {\bm r_{2}}; \cos{\frac{12\pi}{6}}, \frac{5}{2}, \sin{\frac{12\pi}{6} })
	 				,( {\bm r_{3}}; \cos{\frac{4\pi}{6}}, \frac{5}{2}, \sin{\frac{4\pi}{6}} )$ 
	 				& $8$ & $12$ & $4$ &  $+1$ & $-\frac{7\sqrt{3}}{4}\xi $\\
			$ Q_{zx}^{-}$ &\ref{f6}(b)& ${\bm n_{y}}$ 
			&\, $( {\bm r_{1}}; \cos{\frac{4\pi}{6}}, \frac{5}{2}, \sin{\frac{4\pi}{6}} )  
	 				,( {\bm r_{2}}; \cos{\frac{12\pi}{6}}, \frac{5}{2}, \sin{\frac{12\pi}{6} })
	 				,( {\bm r_{3}}; \cos{\frac{8\pi}{6}}, \frac{5}{2}, \sin{\frac{8\pi}{6}} )$ 
	 				& $4$ & $12$ & $8$ &  $-1$ & $-\frac{7\sqrt{3}}{4}\xi $ \\
	 		$ Q_{2}^{0}$ &\ref{f4}(f) & ${\bm n_{z}}$ 
	 		&\, $( {\bm r_{1}}; \cos{\frac{11\pi}{6}}, \sin{\frac{11\pi}{6}},\frac{5}{2} )
	 				,( {\bm r_{2}}; \cos{\frac{3\pi}{6}}, \sin{\frac{3\pi}{6}}, \frac{5}{2} )
	 				,( {\bm r_{3}}; \cos{\frac{7\pi}{6}}, \sin{\frac{7\pi}{6}},\frac{5}{2} )$ 
	 				& $11$ & $3$ & $7$ & $+1$ & $-\frac{7\sqrt{3}}{2}\xi $ \\
	 		$ Q_{2}^{2}$ &\ref{f4}(d)  & ${\bm n_{z}}$ 
	 		&\, $( {\bm r_{1}}; \cos{\frac{7\pi}{6}}, \sin{\frac{7\pi}{6}},\frac{5}{2} )
	 				,( {\bm r_{2}}; \cos{\frac{3\pi}{6}}, \sin{\frac{3\pi}{6}}, \frac{5}{2})
	 				,( {\bm r_{3}}; \cos{\frac{11\pi}{6}}, \sin{\frac{11\pi}{6}},\frac{5}{2} )$ 
	 				& $7$ & $3$ & $11$ & $-1$ & $-\frac{7\sqrt{3}}{2}\xi $ \\
	 		$ -R_{m}$ &\ref{f6}(d) & ${\bm n_{z}}$ 
	 		&\, $( {\bm r_{1}}; \cos{\frac{2\pi}{6}}, \sin{\frac{2\pi}{6}},\frac{5}{2} )
	 				,( {\bm r_{2}}; \cos{\frac{6\pi}{6}}, \sin{\frac{6\pi}{6}},\frac{5}{2} )
	 				,( {\bm r_{3}}; \cos{\frac{10\pi}{6}}, \sin{\frac{10\pi}{6}},\frac{5}{2} )$ 
	 				& $2$ & $6$ & $10$ & $ +1 $ & $\frac{7\sqrt{3}}{2}\xi $ \\
			$ Q_{xy}$ &\ref{f4}(b) & ${\bm n_{z}}$
			&\, $( {\bm r_{1}}; \cos{\frac{10\pi}{6}}, \sin{\frac{10\pi}{6}},\frac{5}{2} )
	 				,( {\bm r_{2}}; \cos{\frac{6\pi}{6}}, \sin{\frac{6\pi}{6}}, \frac{5}{2} )
	 				,( {\bm r_{3}}; \cos{\frac{2\pi}{6}}, \sin{\frac{2\pi}{6}},\frac{5}{2} )$ 
	 				& $10$ & $6$ & $2$ & $-1$ & $-\frac{7\sqrt{3}}{2}\xi $ \\ 
			$ -Q_{yz}^{+}$ & -- & ${\bm n_{x}}$ 
			&\, $( {\bm r_{1}};\frac{5}{2}, \cos{\frac{-\pi}{6}}, \sin{\frac{-\pi}{6}})
	 				,( {\bm r_{2}};\frac{5}{2}, \cos{\frac{-9\pi}{6}}, \sin{\frac{-9\pi}{6}} )
	 				,( {\bm r_{3}};\frac{5}{2}, \cos{\frac{-5\pi}{6}}, \sin{\frac{-5\pi}{6}} )$ 
	 				& $-1$ & $-9$ & $-5$ & $+1$ & $\frac{7\sqrt{3}}{4}\xi $ \\
			$ -Q_{yz}^{-}$ & --& ${\bm n_{x}}$ 
			&\, $( {\bm r_{1}};\frac{5}{2}, \cos{\frac{-5\pi}{6}}, \sin{\frac{-5\pi}{6}} ) 
	 				,( {\bm r_{2}};\frac{5}{2}, \cos{\frac{-9\pi}{6}}, \sin{\frac{-9\pi}{6}} )
	 				,( {\bm r_{3}};\frac{5}{2}, \cos{\frac{-\pi}{6}}, \sin{\frac{-\pi}{6}} )$ 
	 				& $-5$ & $-9$ & $-1$ & $-1$ & $\frac{7\sqrt{3}}{4}\xi $ \\	
	 		$ -Q_{zx}^{+}$ & --& ${\bm n_{y}}$ 
	 		&\, $( {\bm r_{1}}; \cos{\frac{-10\pi}{6}}, \frac{5}{2}, \sin{\frac{-10\pi}{6}} )
	 				,( {\bm r_{2}}; \cos{\frac{-6\pi}{6}}, \frac{5}{2}, \sin{\frac{-6\pi}{6}} )
	 				,({\bm r_{3}}; \cos{\frac{-2\pi}{6}}, \frac{5}{2}, \sin{\frac{-2\pi}{6}} )$ 
	 				& $-10$ & $-6$ & $-2$ & $ +1 $ & $\frac{7\sqrt{3}}{4}\xi $ \\	
			$ -Q_{zx}^{-}$ & --& ${\bm n_{y}}$ 
			&\, $( {\bm r_{1}}; \cos{\frac{-2\pi}{6}}, \frac{5}{2}, \sin{\frac{-2\pi}{6}} )  
	 				,( {\bm r_{2}}; \cos{\frac{-6\pi}{6}}, \frac{5}{2}, \sin{\frac{-6\pi}{6}} )
	 				,( {\bm r_{3}}; \cos{\frac{-10\pi}{6}}, \frac{5}{2}, \sin{\frac{-10\pi}{6}} )$ 
	 				& $ -2$ & $-6$ & $-10$ & $ -1 $  & $\frac{7\sqrt{3}}{4}\xi $  \\
	 		$ -Q_{2}^{0}$ & \ref{f5}(b)& ${\bm n_{z}}$ 
	 		&\, $( {\bm r_{1}}; \cos{\frac{-7\pi}{6}}, \sin{\frac{-7\pi}{6}},\frac{5}{2} )
	 				,( {\bm r_{2}}; \cos{\frac{-3\pi}{6}}, \sin{\frac{-3\pi}{6}}, \frac{5}{2} )
	 				,({\bm r_{3}}; \cos{\frac{-11\pi}{6}}, \sin{\frac{-11\pi}{6}},\frac{5}{2} )$ 
	 				& $-7$ & $-3$ & $-11$ & $+1$ & $\frac{7\sqrt{3}}{2}\xi $ \\
	 		$ -Q_{2}^{2}$ & --& ${\bm n_{z}}$  
	 		&\, $( {\bm r_{1}}; \cos{\frac{-11\pi}{6}}, \sin{\frac{-11\pi}{6}},\frac{5}{2} )
	 				,( {\bm r_{2}}; \cos{\frac{-3\pi}{6}}, \sin{\frac{-3\pi}{6}}, \frac{5}{2})
	 				,( {\bm r_{3}}; \cos{\frac{-7\pi}{6}}, \sin{\frac{-7\pi}{6}},\frac{5}{2} )$ 
	 				& $-11$ & $-3$ & $-7$ & $-1$ & $\frac{7\sqrt{3}}{2}\xi $ \\
	 		$ R_{m}$ &\ref{f6}(c) & ${\bm n_{z}}$ 
	 		&\, $( {\bm r_{1}}; \cos{\frac{-4\pi}{6}}, \sin{\frac{-4\pi}{6}},\frac{5}{2} )
	 				+( {\bm r_{2}}; \cos{\frac{0\pi}{6}}, \sin{\frac{0\pi}{6}},\frac{5}{2} )
	 				+( {\bm r_{3}}; \cos{\frac{-8\pi}{6}}, \sin{\frac{-8\pi}{6}},\frac{5}{2} )$ 
	 				& $-4$ & $0$ & $-8$ & $ +1 $ & $-\frac{7\sqrt{3}}{2}\xi $ \\
			$ -Q_{xy}$ & -- & ${\bm n_{z}}$ 
			&\,  $( {\bm r_{1}}; \cos{\frac{-8\pi}{6}}, \sin{\frac{-8\pi}{6}},\frac{5}{2} )
	 				,( {\bm r_{2}}; \cos{\frac{0\pi}{6}}, \sin{\frac{0\pi}{6}}, \frac{5}{2} )
	 				,( {\bm r_{3}}; \cos{\frac{-4\pi}{6}}, \sin{\frac{-4\pi}{6}},\frac{5}{2} )$ 
	 				& $-8$ & $0$ & $-4$ & $-1$ & $\frac{7\sqrt{3}}{2}\xi $ \\ 
			\hline
			\end{tabular}
			\label{Tab2}
		\end{table*}	

	\renewcommand{\arraystretch}{1}

	\subsection{MQ and rotational interactions of the RKKY mechanism, and Hamiltonian} 

	\begin{figure}[t]
		\includegraphics[width=8cm]{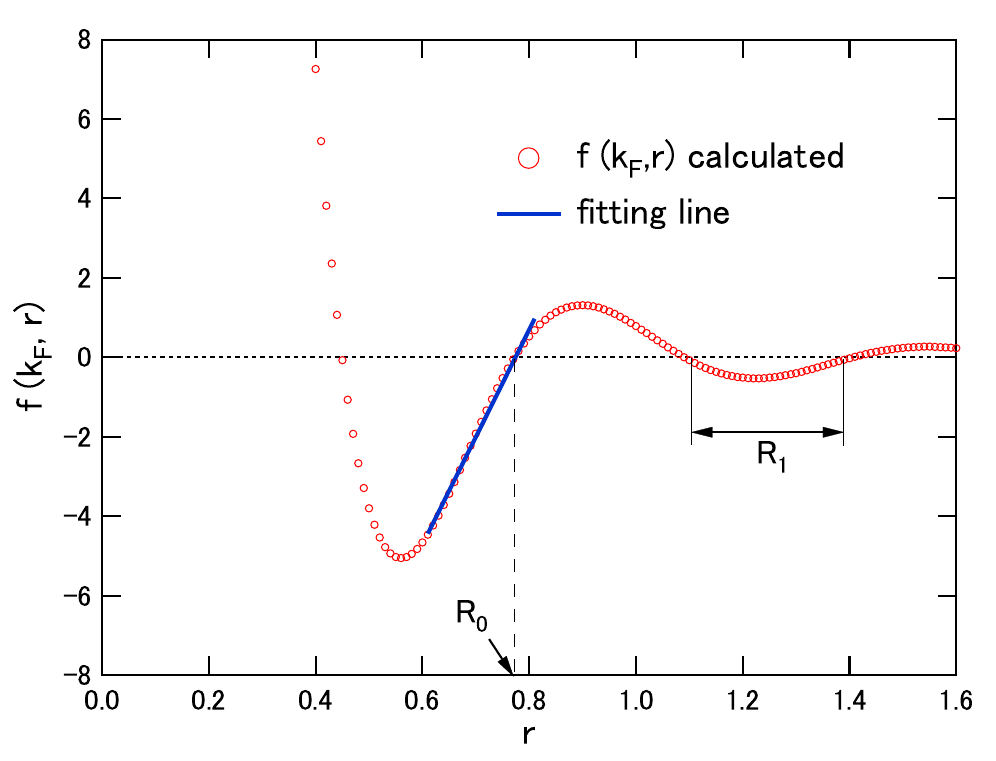}	
		\caption{The open red circles indicate the function $f(k_{\rm F},r)$ calculated by Eq.~\ref{s3_e35bB}. 
		Here, we set the parameter $k_{\rm F}=5$. $R_{0}$ is the second zero point of  $f(k_{\rm F},r)$.
		The solid blue line is the fit to the formula $C_{1}r+C_{2}$ (Eq.~\ref{s3_e37Dd} ). 
		The parameters obtained are $C_{1}=27$, and $C_{2}=-21$.
		The length of the blue line indicates the fitting range. The double-ended arrow ${\rm R_{1}}$ denotes 
		the second range where $f(k_{\rm F},r)$ becomes to be negative.
		}
		\label{f9}
	\end{figure}

In the case of Gd$_{3}$Ru$_{4}$Al$_{12}$, the trimers can be regarded as magnetic impurities with long 
diameter. As shown in Table~\ref{Tab1}, magnitude of $\tilde{Q}_{\Gamma,\gamma}$ and $\tilde{R}_{m}$ are 
proportional to $D$.
Therefore, strength of MQ and rotational interactions are expected to be proportional to $D^{2}$.
Under this special condition in the breathing kagome structure, we derive MQ interactions synthesizing dipole 
interactions of the RKKY mechanism. 
According to the this mechanism, the spin polarization ${\bm p}({\bm r})$ on the Fermi surface at position ${\bm r}$ 
generated by spin $\bm S$ is described as
\cite{Nagamiya}
		\begin{align}
			{\bm p}({\bm r})
			&=\frac{3}{16\pi}VNj_{0}{\bm S}\,\frac{1}{E_{\rm F}}
				f(k_{\rm F},r)  \label{s3_e33gG} \\ 
			&=C_{p} f(k_{\rm F},r) {\bm S}. \label{s3_e34aA}
		\end{align}
Here, $V$ is the volume, $N$ the number of conduction electrons in the Fermi sphere, $E_{\rm F}$ the
Fermi energy, ${\bm S}$ the spin placed at the origin  and constant value $j_{0}$ the Fourier coefficient
of $j({\bm r})$ of $s$--$f$ interaction (Appendix C).
In Eqs.~\ref{s3_e33gG} and \ref{s3_e34aA}, $f(k_{\rm F},r)$ is defined as 
		\begin{align}
			f(k_{\rm F},r)
			=-\left[\frac{\cos({2k_{\rm F}r})}{r^{3}}-\frac{\sin({2k_{\rm F}r})}{2k_{\rm F}r^{4}}\right], 	
			\label{s3_e35bB}
		\end{align}
and $C_{p}$ is the redefined constant
		\begin{align}
			C_{p}=\frac{3}{16\pi}VNj_{0}\,E_{\rm F}^{-1}.
			\label{s3_e36cC}
		\end{align}

The open red circles in Fig.~\ref{f9} denote the $f(k_{\rm F},r)$ as a function of $r$ by Eq.~\ref{s3_e35bB}
with the parameter $k_{\rm F}=5$. In this figure, $R_{0}$ is the second zero point of $f(k_{\rm F},r)$. The solid
blue line in Fig.~\ref{f9} is the fit to the formula
		\begin{align}
			f(k_{\rm F},r)
			&=C_{1}r + C_{2}  \label{s3_e37Dd} \\
			&= C_{1}\Delta l 
			\label{s3_e38Ee}
		\end{align}
where $C_{1}$ and $C_{2}$ are constants, and $\Delta l$ is defined as the distance between $r$ and $R_{0}$,
		\begin{align}
			 \Delta l=r - R_{0}.
			 		\label{s3_e39Dd}	
		\end{align}
In the previous study, 
it was pointed out that the transition from the FM to the AFM behavior of Gd$_{3}$Ru$_{4}$Al$_{12}$ with decreasing 
temperature is interpreted as results of synthesis of dipole interactions of the RKKY mechanism
\cite{Nakamura2018}. 
Because the distances of two spins which respectively belongs to trimers next to each other are
approximately equal to $R_{0}$, magnetic properties of Gd$_{3}$Ru$_{4}$Al$_{12}$ are determined by delicate
competition of the FM and AFM interactions. 
From this point of view, we approximate $f(k_{\rm F},r)$ in Eq.~\ref{s3_e35bB} by Eqs.~\ref{s3_e38Ee} and \ref{s3_e39Dd}.

\begin{figure}[b]
		\includegraphics[width=8.5cm]{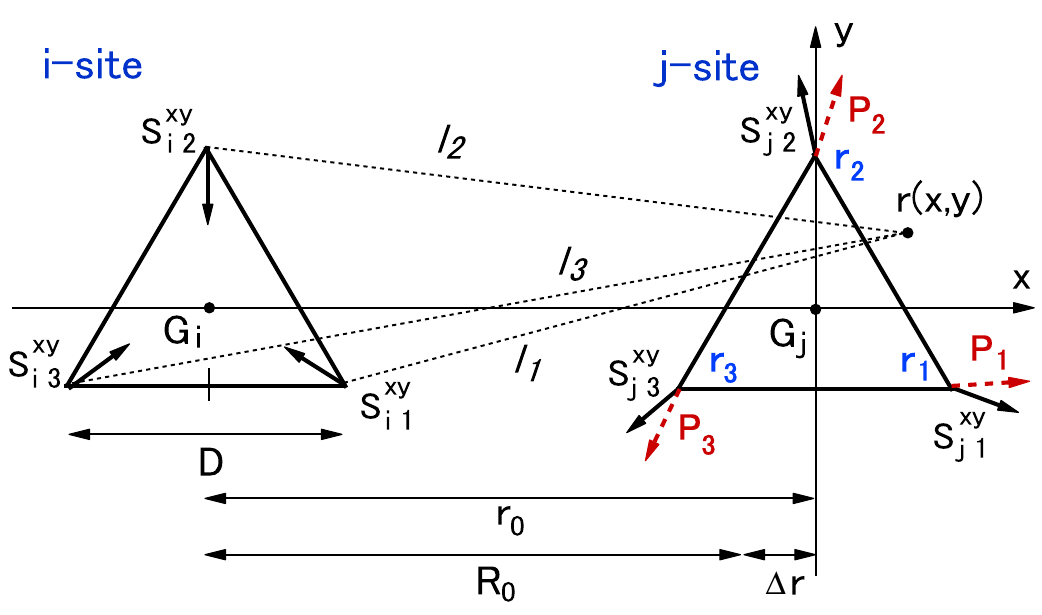}	
		\caption{Induced spin polarization at ${\bm r}(x,y)$ by the RKKY mechanism. 
		The broken-red arrows ${\bm p_{j}}$ $(j=1,2,3)$ are spin polarization at ${\bm r_{j}}$ ($j=1,2,3$) induced by 
		${\bm S_{ik}^{xy}}$ ($k=1,2,3$).
		Here, ${\rm G_{i}}$ and ${\rm G_{j}}$
		are the centers of gravity of the triangles, and ${\rm G_{j}}$ is placed on the origin. 
		$r_{0}$ is the distance between ${\rm G_{i}}$ and ${\rm G_{j}}$. $D$ denotes the 
		length of the sides of the triangle, and ${\it l_{k}}$ $(k=1,2,3)$ are the distances 
		between ${\bm S_{ik}}$ and ${\bm r}(x,y)$, respectively. 
		}
		\label{f10}
	\end{figure}

In Fig.~\ref{f10}, two trimers on $i$ and $j$-sites are illustrated. Three $xy$ plane component spins ${\bm S_{ik}^{xy}}$ $(k=1,2,3)$
are placed on the vertexes of the trimer on $i$-site, and ${\bm S_{jk}^{xy}}$ $(k=1,2,3)$ are placed on $j$-site. 
In Fig.~\ref{f10}, $r_{0}$ is the distance between the centers of the gravity of triangles next to each other. 
Here, we define $\Delta r$ as
		\begin{align}
			 \Delta r=r_{0}-R_{0}.
			 		\label{s3_e40ee}	
		\end{align}
 From assumptions we mentioned before, $\Delta r$ is a short length.

 Using Eqs.~\ref{s3_e34aA}, \ref{s3_e38Ee} and \ref{s3_e40ee},
  spin polarization induced by the trimer on $i$-site  at ${\bm r_{1}}$ in Fig.~\ref{f10} is given by

		\begin{align}
			{\bm p}({\bm r_{1}})
			&=\sum_{k=1,2,3}C_{p} f(k_{\rm F},l_{k}) {\bm S_{ik}^{\alpha}} \label{s3_e41Ff} \\
			&\simeq C_{p}C_{1}\bigg[ -\frac{D^{2}}{2R_{0}} {\bm S_{i1}^{\alpha}} +\frac{r_{0}D}{2R_{0}} 
			 ({\bm S_{i3}^{\alpha}} -  {\bm S_{i1}^{\alpha}})\bigg]
			 		\label{s3_e42gG}		 
		\end{align}
(Appendix D).
Similarly, spin polarization at ${\bm r_{2}}$ and at ${\bm r_{3}}$ induced by trimer A are described as
		\begin{align}
			{\bm p}({\bm r_{2}})
			&\simeq C_{p}C_{1}\bigg[ -\frac{D^{2}}{2R_{0}} {\bm S_{i2}^{\alpha}} +\frac{r_{0}D}{2R_{0}} 
			 ({\bm S_{i3}^{\alpha}} -  {\bm S_{i1}^{\alpha}})\bigg],
			 		\label{s3_e43aAA} 
		\end{align}
		\begin{align}		 
			{\bm p}({\bm r_{3}})
			&\simeq C_{p}C_{1}\bigg[ -\frac{D^{2}}{2R_{0}} {\bm S_{i3}^{\alpha}} +\frac{r_{0}D}{2R_{0}} 
			 ({\bm S_{i3}^{\alpha}} -  {\bm S_{i1}^{\alpha}})\bigg],
			 		\label{s3_e44bBB}		 
		\end{align}
respectively.	
These spin polarization do not affect the basic FM dipole features of the trimer 
because the spin polarization in Eqs.~\ref{s3_e42gG}-\ref{s3_e44bBB} are perpendicular to the directions of ${\bm S_{r}^{n}}$.
In the process of calculation we are conscious of the relational formula
		\begin{align}
			{\bm S_{j1}^{\alpha}}+{\bm S_{j2}^{\alpha}}+{\bm S_{j3}^{\alpha}}={\bm 0}.
			 		\label{s3_e45cC}	 
		\end{align}

Thus, the MQ interactions between the trimers on $i$ and $j$-sites are described as
		\begin{align}
			-\sum_{\Gamma \gamma,\sigma} G_{\Gamma}\,  
			Q_{\Gamma \gamma}^{\sigma}(i)  Q_{\Gamma \gamma}^{\sigma}(j).
			 		\label{s3_e46dDD}		 
		\end{align}
Here, $G_{\Gamma}$ are the coupling constants whose dimensions are energy, $\sigma$ indicates the sign of chirality. 
When there are different QS which possesses different chiralities,
the summation runs over $\sigma$ also. MQ moments belong to different chiralities do not
interact with each other.
Similarly to MQ interactions, the rotational interaction is given as
		\begin{align}
			-G_{R}\,R_{m}(i) R_{m}(j),
			 		\label{s3_e47bB}		 
		\end{align}
where $G_{R}$ is the coupling constant.
In Eq.~\ref{s3_e46dDD}, coupling constants are defined as
		\begin{align}
			G_{\Gamma\gamma}
			&\propto - C_{p} C_{1} \frac{D^{2}}{R_{0}^{2}} \left(2j_{0} \right) \label{s3_e48cCC}\\
			&=- C_{1}\frac{3}{8\pi}VNj_{0}^{2}\frac{1}{E_{\rm F}}\frac{1}{R_{0}^{2}}<0.
			 		\label{s3_e49DDd}		 
		\end{align}
Because $C_{1}$ shows a positive value, the MQ interaction act on the trimers in Fig.~\ref{f10}
is AFM. The coupling constant $G_{R}$ (Eq.~\ref{s3_e47bB}) is also negative.

Adding to the MQ interactions, magnetic dipole interactions acts between ${\bm S_{r}}$s.
Therefore, Hamiltonian in the low temperatures is described as follows.
	\begin{widetext}
		\begin{align}
			\mathscr{H}_{0}= &-\frac{1}{2}J_{1}\sum_{{\rm NN}(i,j)}  {\bm S_{r}(i)} {\bm S_{r}(j)}
			- \frac{1}{2}\sum_{{\rm NN}(i,j)} \biggl[\,\, \sum_{\Gamma \gamma,\sigma} G_{\Gamma}\,  
			 Q_{\Gamma \gamma}^{\sigma}(i)  Q_{\Gamma \gamma}^{\sigma}(j)
				+  G_{R}\,R_{m}(i) R_{m}(j)\biggr] \notag\\
			&-\frac{1}{2}J_{2}\sum_{{\rm NNN}(k,l)}  {\bm S_{r}(k)} {\bm S_{r}(l)} 
			-\frac{1}{2}J_{3}\sum_{{\rm 3dNN}(m,n)}  {\bm S_{r}(m)} {\bm S_{r}(n)}
				+\mathscr{H}_{Z}.
			 		\label{s3_e50eHH}		 
		\end{align}
	\end{widetext}
Here, the symbols $J_{1}$, $G_{\Gamma}$, $G_{R}$, $J_{2}$ and $J_{3}$ are coupling constants, and
the characters $i$, $j$, $k$ and $l$ in the parentheses denote trimer sites.
The symbols ${\rm NN}$, ${\rm NNN}$ and ${\rm 3dNN}$ in the summations mean to take sum over the nearest
neighbor (NN), next nearest neighbor (NNN) and the third nearest neighbor (3dNN) trimers in the same Gd--Al
layer, respectively. The coefficient (1/2) applied to each term is for elimination of duplication.
The symbol $\sigma$ on the right shoulder of $Q$ denote signs of chiralities. 
When  different chirality of $Q_{\Gamma\gamma}$ exist, the third summation also runs over $\sigma$.
Even if symmetry $\Gamma\gamma$ of the QSs of two trimers are the same, they do not
interact with each other when their signs of chiralities are different. For example, $Q_{yz}^{+}$ and $Q_{yz}^{-}$ do not
interact with each other. Globally, chiral domains may be generated.
The last term in the right side of Eq.~\ref{s3_e50eHH} is Zemann term determined
by $[{\bm S_{r}}(1),{\bm S_{r}}(2),...]$.
We neglect quantum kinetic energy in Eq.~\ref{s3_e50eHH} because we regarded the spins as classical spins.
Because reaching distance of the interactions between such high rank moments are not expected to be long,
we consider MQ interactions between NN trimers only. On the other hand, reaching distance of dipole interaction 
would be longer compared to those of MQ interactions, then, we consider to the third nearest neighbor trimers.
Because the distance between NN trimers is approximately equal to $R_{0}$ in Fig.~\ref{f9}, magnitude of the
first term in the right side of Eq.~\ref{s3_e50eHH} is small, i.e.,
		\begin{align}
			|J_{1}| \ll |J_{2}|,|J_{3}|.
			 		\label{s3_e51fFF}		 
		\end{align}
If we assume the isotropy in the $ab$ plane,
		\begin{align}
			G_{R} = G_{\Gamma 1}
			 		\label{s3_e52fGG}.		 
		\end{align}
We assume that the both distances between NNN trimers and 3dNN trimers are in the range $R_{1}$ in Fig.~\ref{f9}
where $f(k_{\rm F},l_{i})$ is negative. According to this assumption, both $J_{2}$ $J_{3}$ are negative.
Actually, Gd$_{3}$Ru$_{4}$Al$_{12}$ shows AFM behaviors at low temperatures as represented in Fig.~\ref{f8}.

\section{Magnetic phase transitions and magnetic anisotropies}

	\subsection{Separation of two sub spin-systems and the partial order} 

In this section, we examine the consistency between the Hamiltonian and macroscopic
magnetic properties actually have been observed.
After this, we comprehensively represent the coupling constants $G_{\Gamma n}$ as $G_{n}$ and
QSs which belongs to $\Gamma_{n}$ symmetry as $Q_{n}$ for easy viewing depending on the case.
Some discussion about quadrupole frustration, partial orders and coexistence of different types of magnetic anisotropies
in Gd$_{3}$Ru$_{4}$Al$_{12}$ have been made by Nakamura {\it et. al}
\cite{Nakamura2023}.
They point out that MQ interactions play important roles to express these magnetic properties.
However, their discussion is not based on the microscopic mechanisms of interactions.

\begin{figure}[t]
		\includegraphics[width=8.5cm]{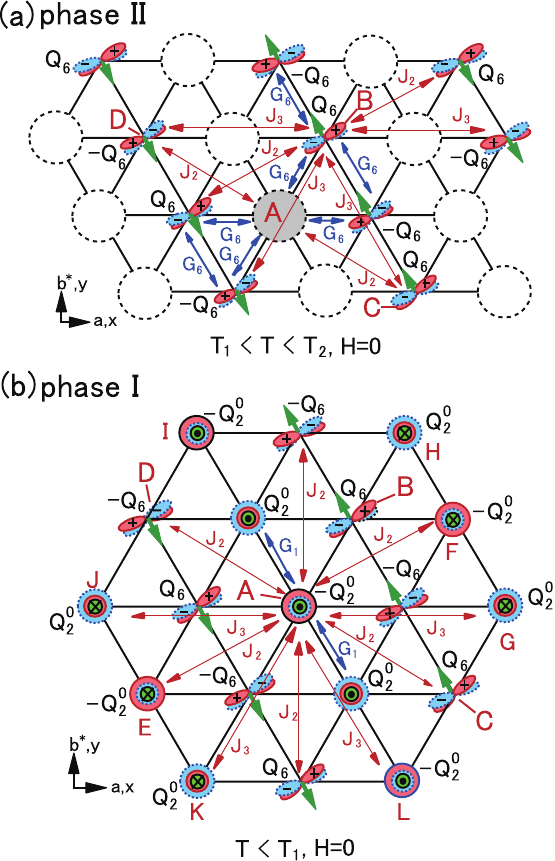}	
		\caption{The structures of ${\bm S_{r}}$ ($S_{r}=15/2$) of Gd$_{3}$Ru$_{4}$Al$_{12}$
\cite{Matsumura2019}		
		and the structure of QSs assumed in the present study 
		 in (a) phase II and (b) phase I  at zero field. In this figure, the coordinate $x-y-z$ is defined
		 as $x\parallel a$, $y\parallel b^{*}$ and $z\parallel c$.
		 The green arrows indicate ${\bm S_{r}}$s. The butterfly shapes denote 
		  $Q_{6}$s, and the double circle indicates $Q_{2}^{0}$s. 
		According to the previous paper, $T_{1}=17.5$ K and $T_{2}=18.6$ K
		have been obtained
\cite{Nakamura2018}.
		The broken black circles in Fig.~\ref{f11}(a) denote disordered trimer sites.
		The red and blue arrows indicate interactions in Eq.~\ref{s3_e50eHH}.	
		Not all of the interactions are illustrated for ease of viewing. The green $\odot$ and $\otimes$ denote ${\bm S_{r}}$s
		directed along and against the $c$ axis, respectively.
		}
		\label{f11}
	\end{figure}

As we mentioned before, the trimers form a triangular lattice on the breathing kagome lattice of Gd$_{3}$Ru$_{4}$Al$_{12}$.
Figure~\ref{f11} shows the structure of ${\bm S_{r}}$ at zero field determined by the RXD measurements
\cite{Matsumura2019}.
As illustrated in Fig.~\ref{f11}, phase II is the partial ordered IMT phase which appears in the range $T_{1} <T < T_{2}$, 
and phase I is the fully 
ordered low temperature phase which appears in the range $T<T_{1}$.
The bold green arrows in Fig.~\ref{f11} indicate ${\bm S_{r}}$s.
The directions of ${\bm S_{r}}$s in Fig.~\ref{f11}(a) are parallel to the $b$ axis, and the broken circles
denote disordered trimer sites.
Overposing the ${\bm S_{r}}$s structure in Fig.~\ref{f11}(a), we illustrate a structure of 		
QSs with butterfly shapes. 
The symbol $Q_{6}$ in Fig.~\ref{f11} indicates a linear combination of $Q_{zx}^{+}$ and $Q_{yz}^{+}$ described as
		\begin{align}
			Q_{6}^{+}\left(\frac{\pi}{6}\right) = \left( \cos{\frac{\pi}{6}} \right) Q_{zx}^{+} 
						+ \left( \sin{\frac{\pi}{6}} \right) Q_{yz}^{+}
			 		\label{s4_e53H}		 
		\end{align}
 (Appendix E). When we assume this linear combination of two QSs, ${\bm S_{r}}$ is directed parallel to the $b$ axis.
The blue bi-directional arrows in Fig.~\ref{f11} with symbol ``$G_{6}$'' indicate MQ interactions between NN,
and the red arrows with ``$J_{2}$''
and ``$J_{3}$'' denote dipole interactions between NNN and 3dNN trimers, respectively. 
These labels correspond to the coupling constants in Eq.~\ref{s3_e50eHH}.
As we mentioned before, we neglect the dipole interactions between
NN trimers. For the sake of easy look, not all of interactions are illustrated in Fig.~\ref{f11}.

Here, we focus on the disordered trimer site ``A'' in Fig.~\ref{f11}(a) which is painted with light gray.
This trimer receives AFM MQ interactions from the four surrounding $\pm Q_{6}$s as shown by the blue bi-directional
arrows. The trimer on A site is surrounded by six NN trimers, however, two of them are disordered sites.
The MQ interactions generated by the remaining four $\pm Q_{6}$s
are canceled with each other on A site, therefore, site A becomes a quadrupole
disordered site in the range $T_{1} <T < T_{2}$. 
The trimer on A site also does not receive dipole interactions from the six surrounding NN 
trimers. Adding this, dipole interactions from the surrounding four NNN trimer sites with spontaneous dipole moments 
cancel out with each other. 
For example, interactions between A--D sites and those between A--C sites in Fig.~\ref{f11}(a) cancel out with each 
other on A site.
Further, all of six 3dNN sites surround A site are disordered sites.  Therefore, the trimer on A site does not receive
interactions from any spontaneous moments illustrated in Fig{f11}(a).

Next, we focus on ``B'' site in Fig.~\ref{f11}(a).
The trimer on B site receives MQ interactions $G_{6}$ from two trimers on the NN sites. This induces $Q_{6}$ on B site.
As shown in Fig.~\ref{f11}(a), the trimer on B site receives AFM dipole interactions $J_{3}$ from two trimers on 3dNN sites
and receives AFM dipole interactions $J_{2}$ from two trimers on NNN sites. Other interactions from 3dNN and NNN sites
are canceled out with each other. This leads to the generation of spontaneous dipole moments on B site.
The change in energy originating from the dipole interactions related to B site is given as
		\begin{align}
			2J_{2}\left(S_{r}\right)^{2}+2J_{3}\left(S_{r}\right)^{2},
			 		\label{s4_e54Ii}		 
		\end{align}
where $S_{r}=15/2$ (Appendix B). Thus, the change in the energy per trimer
 which is placed on the equivalent site of B in phase II originating from dipole interactions is
		\begin{align}
			E_{d}^{B}=J_{2}\left(S_{r}\right)^{2}+J_{3}\left(S_{r}\right)^{2}.
			 		\label{s4_e55jJj}		 
		\end{align}
Similarly, the change in the energy per trimer originating from MQ interactions on the equivalent site of B is
		\begin{align}
			E_{q}^{B}=G_{6}\left(Q_{6}\right)^{2}.
			 		\label{s4_e56KKk}		 
		\end{align}
Therefore, the total change in the energy originating from multipole interactions is described as
		\begin{align}
			E_{mp}^{B}=G_{6}\left(Q_{6}\right)^{2}+J_{2}\left(S_{r}\right)^{2}+J_{3}\left(S_{r}\right)^{2}
			 		\label{s4_e57LlL}		 
		\end{align}
per trimer site which is equivalent of B site.

We consider the structures of ${\bm S_{r}}$s and QSs in phase I at zero field.
As shown in Fig.~\ref{f11}(b), we assume the spontaneous $-Q_{2}^{0}$ 
appears on A site (the center of the hexagon). Here, A, B, C sites in Fig.~\ref{f11}(b) are the same sites in Fig.~\ref{f11}(a).
In Fig.~\ref{f11}(b), both chiralities of $\pm Q_{2}^{0}$ are plus as shown in Table~\ref{Tab2}.
The $-Q_{2}^{0}$ on A site in this figure does not interact with the surrounding four 
$\pm Q_{6}$s on NN sites because of mismatch of the symmetries. 
On the other hand, $-Q_{2}^{0}$ on A site interact with two $Q_{2}^{0}$s on remaining NN sites. 
Adding this, the trimer on the A site in Fig.~\ref{f11}(b) interacts with surrounding four NNN trimers with $\pm Q_{6}$
in the dipole manner. However, these dipole interactions are canceled out with each other.
For example, dipole interactions from the trimers on C and D sites in Fig.~\ref{f11}(b) are canceled with each other on A site.
On the other hand, dipole interactions with the remaining two NNN trimers (on E and F sites) with $-Q_{2}^{0}$ 
in this figure are not canceled.
The trimer on A site in Fig.~{f11}(b) interacts with six 3dNN trimers on G--L sites with dipole manner, 
but the interactions with four of them are canceled.
Consequently, the change in energy per trimer originating from the multipole interactions 
on the equivalent site of A is obtained as
		\begin{align}
			E_{mp}^{A}=G_{1}\left(Q_{2}^{0}\right)^{2}+J_{2}\left(S_{r}\right)^{2}+J_{3}\left(S_{r}\right)^{2}.
			 		\label{s4_e58MM}		 
		\end{align}

Thus, the trimers in Fig.~\ref{f11}(b) separates into two groups which do not interact with each other. 
One group includes trimers
which possesses $\pm Q_{6}\left(\frac{\pi}{6}\right) $ and the other includes trimers which possesses $\pm Q_{2}^{0}$. 
Suppressing frustration, the system gets energy gain deduced from dipole and quadrupole interactions.
These long range orders of two subgroups would disappear at different temperatures with increasing temperature.

	\begin{figure}[b]
		\includegraphics[width=8.5cm]{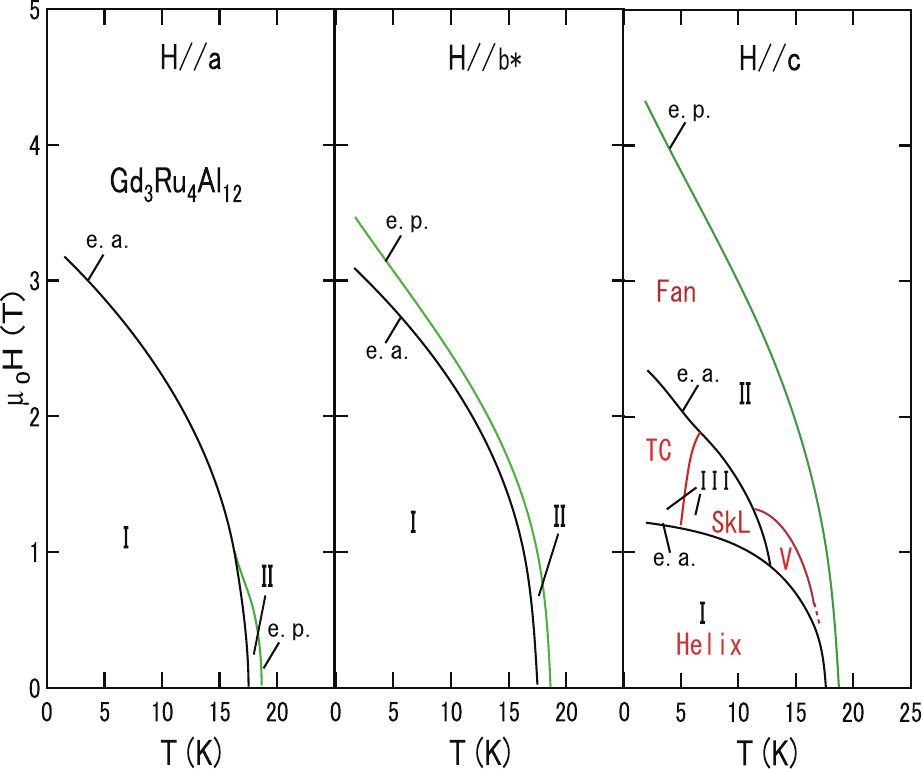}	
		\caption{Schematic phase diagrams of Gd$_{3}$Ru$_{4}$Al$_{12}$ based on previous studies.
		The phase boundaries illustrated with black and green curves are observed
		in macroscopic measurements
		\cite{Nakamura2023}, and
		red curves denote phase boundaries only observed in RXD
		\cite{Hirschberger2019_2}.
		The names of phases written in black letters are referred from Ref.~5 and those written in red
		letters are referred from Ref.~9. The letters ``SkL'' and ``TC'' denote the skyrmion lattice and 
		transverse conical phases,
		 respectively.
		 The symbols ``e. p.'' and ``e. a.'' indicate easy plane and easy axis magnetic anisotropies.
		}
		\label{f12}
	\end{figure}
%

	\subsection{Magnetic anisotropies generated by the MQ interactions} 

\begin{figure}[b]
		\includegraphics[width=8.5cm]{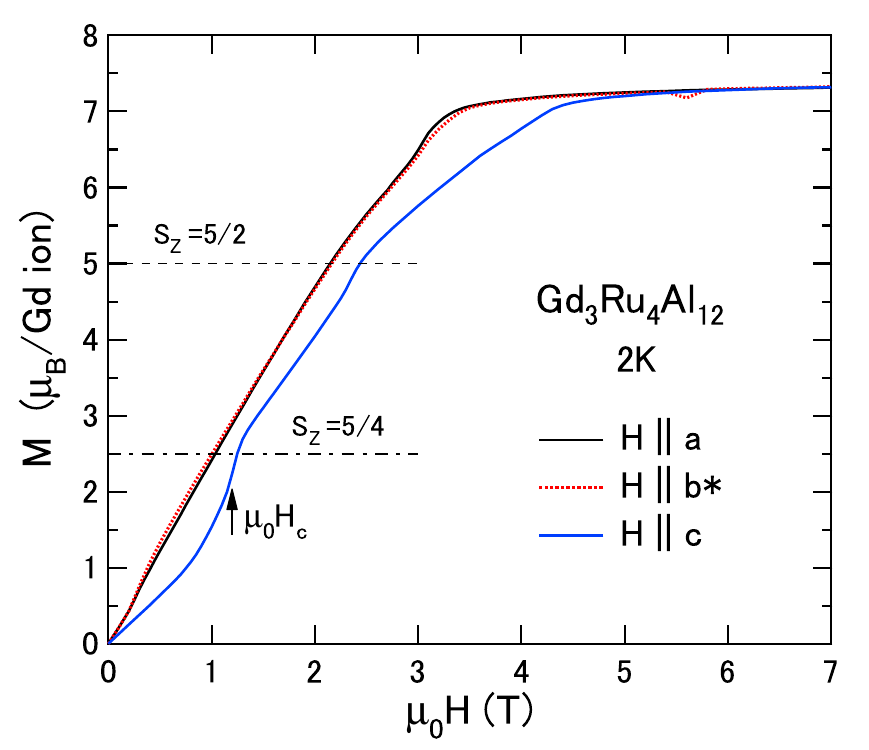}	
		\caption{The magnetization process of Gd$_{3}$Ru$_{4}$Al$_{12}$ at 2 K.
		Data is taken from the previous paper
		\cite{Nakamura2023}. 
		Magnetic fields are directed along the $a$, $b^{*}$ and $c$ axes.
		The arrow indicates the transition field ($\mu_{0}H_{c}$) from the phase I to the phase III. This phase transition
		is associated with the jump in the magnetization. The horizontal broken line, 
		dotted broken line indicate magnetization corresponds to $S_{z}=5/2$, $S_{z}=5/4$, 
		respectively.
		}
		\label{f13}
	\end{figure}

We present schematic phase diagrams of Gd$_{3}$Ru$_{4}$Al$_{12}$ in Fig.~\ref{f12}. 
The phase boundaries illustrated with black and green curves in Fig.~\ref{f12} are determined by macroscopic measurements
\cite{Nakamura2023},
and the boundaries represented with the red curves are phase boundaries observed in RXD measurements
\cite{Hirschberger2019_2}.
No clear anomalies are observed in the specific heat on the boundaries represented with red curves, therefore, 
nature of the phase transitions on these red curves may be changes in long period spin (${\bm S_{r}}$) structures. 
According to the RXD measurements, phase III determined by macroscopic measurements further separates
into transverse conical (TC) phase and magnetic SkL phase. 
Although 4$f$ electrons of Gd$^{3+}$ possesses no angular momentum, the magnetic phase diagrams
show clear magnetic anisotropies as shown in Fig.~\ref{f12}.
It has been pointed out that the phase diagrams of Gd$_{3}$Ru$_{4}$Al$_{12}$ looks like an 
overlapping of two sets of magnetic phase diagrams which show easy plane anisotropy and which show easy axis anisotropy
\cite{Nakamura2023}.
The symbol ``e. p.'' and the green curves in Fig.~\ref{f12} denote the easy plane type magnetic phase diagrams,
and the symbol ``e. a.'' and the black curves indicate easy axis type magnetic phase diagrams.
This implies that the trimers are separated into two subgroups which have different anisotropies actually.
In the easy axis type magnetic anisotropy, the favorable direction for ${\bm S_{r}}$ is parallel to the $c$ axis.
Therefore, when magnetic fields are applied along the $c$ axis, spin flop occurs,
and the jump in the magnetization appears at the phase I/phase III transition field.
The magnetization process of Gd$_{3}$Ru$_{4}$Al$_{12}$ at 2 K is presented in Fig.~\ref{f13}.
As shown in this figure, the jump in the magnetization is observed at the transition field $\mu_{0}H_{c}$ when fields are applied 
along the $c$ axis actually.
On the other hand, e. p. type phase boundaries in Fig.~\ref{f12} are single curves without spin flop.
Corresponding this, no clear jumps in the magnetization are shown in Fig.~\ref{f13} when fields are applied parallel to
 $a$ or $b^{*}$ axes.

	\begin{figure}[t]
		\includegraphics[width=8.5cm]{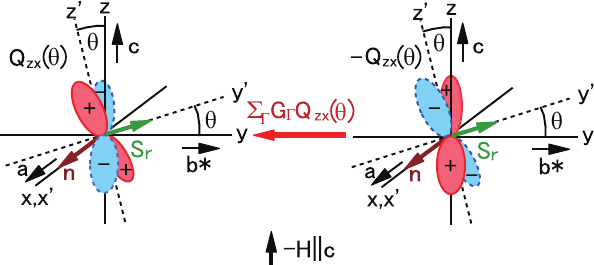}	
		\caption{The QSs $\pm Q_{zx}(\theta)$ placed in the NN trimer sites and interact with each other.
		The green arrows indicate ${\bm S_{r}}$s.
		The $xy$ plane is parallel to the $ab$ plane and the $z$ axis is directed along the $c$ axis.  
		The unit vector ${\bm n}$ is directed along the $x$ axis.
		Each moment is rotated by angle $\theta$ around ${\bm n}$ due to the influence of the weak applied 
		field $-{\bm H}\parallel c$.
		The bold red arrow denotes the effective QS field $\sum_{\Gamma}G_{\Gamma}Q_{zx}(\theta)$
		generated by $-Q_{zx}(\theta)$ and acts on $Q_{zx}(\theta)$.
		}
		\label{f14}
	\end{figure}

Figure~\ref{f14} displays two QSs under the fields directed along the $c$ axis.
 In this figure, $Q_{zx}$ and  $-Q_{zx}$ are rotated around the unit vector ${\bm n}$ by
angle $\theta$ commonly due to the weak applied field $-{\bm H}\parallel c$. 
The relation between the coordinates in Fig.~\ref{f14} is given as
		\begin{align}
				\begin{pmatrix}
					x'			& y'				& z'			
				\end{pmatrix}
				=R_{x}
				\begin{pmatrix}
					x			& y				& z		
				\end{pmatrix}.			
		\label{s4_e59BBB}	 
		\end{align}
Here,
		\begin{align}
			R_{x}=&
				\begin{pmatrix}
					1			&  0				& 0				\\	
					0			&  \cos{\theta}	& -\sin{\theta}	\\
					0			&  \sin{\theta}	&  \cos{\theta}
				\end{pmatrix}.		
		\label{s4_e60cCc}	 
		\end{align}
The QS $Q_{zx}(\theta)$ in Fig.~\ref{f14} is described as functions of $\theta$ in 
the $x$--$y$--$z$ coordinate as
		\begin{align}
			Q_{zx}(\theta)&=Q_{zx} \cos{\theta} -  Q_{xy} \sin{\theta}.	
				\label{s4_e61cDD}	
		\end{align}
Similarly, $-Q_{zx}(\theta)$ in Fig.~\ref{f14} is given as 
		\begin{align}
			-Q_{zx}(\theta)&=-Q_{zx} \cos{\theta} +  Q_{xy} \sin{\theta}.	
				\label{s4_e62eeE}	
		\end{align}
As shown in Fig.~\ref{f14} and according to Eq.~\ref{s4_e62eeE}, $-Q_{zx}(\theta)$ induces the 
additional effective QS fields on $Q_{zx}(\theta)$. The change in the energy of $Q_{zx}(\theta)$ arising from
the quadrupole interaction in Fig.~\ref{f14} is obtained as a function of $\theta$ 
		\begin{align}
			&G_{6} (Q_{zx})^{2}\cos^{2}{\theta} +G_{5} (Q_{xy})^{2}\sin^{2}{\theta} \notag \\
			&=G_{6}(Q_{6})^{2}+\left(4G_{5}-G_{6}\right) (Q_{6})^{2} \sin^{2}{\theta}.			
					\label{s4_e63FfF}	 
		\end{align}
Here, $(Q_{5})^{2}=4(Q_{6})^{2}$ by the definition (Eq.~\ref{s3_e23ff}, Table~\ref{Tab1}).
The first and the second terms on the right side of Eq.~\ref{s4_e63FfF} tend to direct ${\bm S_{r}}$s
in the $ab$ plane under weak external magnetic fields when
		\begin{align}
			0>4G_{5}>G_{6}.	
								\label{s4_e64ggG}	
		\end{align}
	\begin{figure}[t]
		\includegraphics[width=8.5cm]{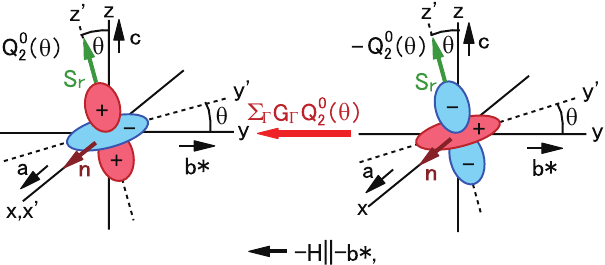}	
		\caption{The QSs $\pm Q_{2}^{0}(\theta)$ placed in the NN and interact with each other.
		The green arrows indicate ${\bm S_{r}}$s.
		The $xy$ plane is parallel to the $ab$ plane and the $z$ axis is directed along the $c$ axis. 		 
		The unit vector ${\bm n}$ is directed along the $x \parallel a$ axis.
		Each moment is rotated by angle $\theta$ around ${\bm n}$ due to the influence of
		 the weak applied field $-{\bm H}\parallel -b^{*}$.
		The bold red arrow  denotes the effective QS field $\sum_{\Gamma}G_{\Gamma}Q_{2}^{0}(\theta)$
		generated by $-Q_{2}^{0}(\theta)$ and acts on $Q_{2}^{0}(\theta)$.
		}
		\label{f15}
	\end{figure}

Figure \ref{f15} shows a pair of  $Q_{2}^{0}(\theta)$ and $-Q_{2}^{0}(\theta)$. 
Each QS 
is rotated around the vector ${\bm n}$ by angle $\theta$ 
due to the magnetic field $-{\bm H}\parallel -b^{*}$. Here, unit vector ${\bm n}$ is directed along the $x$ axis.
Similar to the case of $Q_{6}$, $Q_{2}^{0}$ in Fig.~\ref{f15} is described as functions of $\theta$ in the $x$--$y$--$z$ 
coordinate as
		\begin{align}
			\pm Q_{2}^{0}(\theta)&=\pm Q_{2}^{0}\left(1-\frac{3}{2}\sin^{2}{\theta}\right) 
			\mp Q_{2}^{2} \, \frac{3}{2}\sin^{2}{\theta} \notag \\
			&\quad \mp Q_{yz} \,3 \sin{\theta}\cos{\theta},
			\label{s4_e65CCH}		
		\end{align}
where double sign corresponds.
The change in the energy of $Q_{2}^{0}$ in Fig.~\ref{f15} due to the MQ interaction is given as
		\begin{align}
			&G_{1}(Q_{2}^{0})^{2}+3\sin^{2}{\theta} \left[3G_{6}(Q_{yz})^{2} -G_{1}(Q_{2}^{0})^{2}\right]   \label{s4_e66Ddd}	\\
			=& G_{1}(Q_{2}^{0})^{2} + 3\sin^{2}{\theta} \left(3G_{6}-4G_{1}\right) (Q_{6})^{2}.
			\label{s4_e67Eee}		
		\end{align}
Here, we ignore higher order terms of $\sin^{4}{\theta}$.
When the sign of  $3G_{6}-4G_{1}$ in Eq.~\ref{s4_e67Eee} is positive, ${\bm S_{r}}$s tend to direct along the $c$ axis. 
Therefore,
		\begin{align}
			0>G_{6}>\frac{4}{3}G_{1}.	
								\label{s4_e68fFf}	
		\end{align}
According to Eqs.~\ref{s4_e64ggG} and \ref{s4_e68fFf},  the condition of the coexistence of the easy plane and 
easy axis magnetic anisotropies indicated by the magnetic phase diagrams is obtained as
		\begin{align}
			0>4G_{5}>G_{6}>\frac{4}{3}G_{1}.	
								\label{s4_e69gGg}	
		\end{align}
	\begin{figure}[t]
		\includegraphics[width=8.5cm]{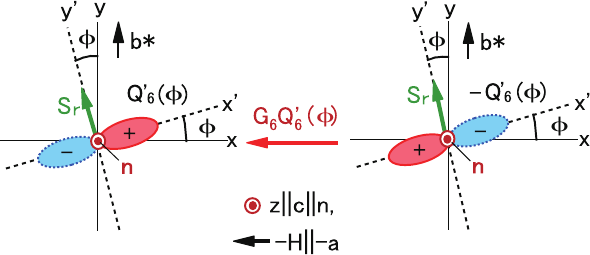}	
		\caption{The QSs $Q_{6}'(\phi)$ and $-Q_{6}'(\phi)$ which placed in the NN under applied field ${-\bm H}$.
		The $xy$ plane is parallel to the $ab$ plane, and the $z$ is parallel to the $c$ axis. 
		${-\bm H}$ are parallel to the $a$ axis, and ${\bm S_{r}}$s are rotated from $b^{*}$ axis by angle $\phi$ under
		the field. The solid red arrows denotes the QS fields.
		}
		\label{f16}
	\end{figure}

As we mentioned above, the coexistence of the easy axis and easy plane types anisotropies are explained
in terms of MQ interactions. 
In Fig.~\ref{f16}, a pair of QSs $\pm Q'_{6}(\phi)$ are illustrated. When field (${\bm -H}\parallel -a$) are applied, 
${\bm S_{r}}$ is
rotated around ${\bm n}$ ($\parallel c$) by the angle $\phi$, the change in the energy arising from MQ interactions 
is given as
		\begin{align}
			G_{6}(Q'_{6})^{2} \left(\cos^{2}{\phi}+\sin^{2}{\phi} \right)=G_{6}(Q'_{6})^{2}
								\label{s4_e70gGgGG}	
		\end{align}
Therefore, the MQ interaction does not induce magnetic anisotropies in the $ab$ plane. Actually, the magnetic susceptibility
is isotropic in the $ab$ plane as shown in Fig.~\ref{f8}.

	\subsection{The helical--TC phase transition} 

As shown in Fig.~\ref{f12}, the structures of ${\bm S_{r}}$s translates from Helix phase to TC phase
when fields are applied along the $c$ axis
\cite{Matsumura2019,Hirschberger2019_2}.
These phases are adjacent to SkL phase.
According to Monte Carlo simulation on three dimensional chiral magnet such as MnSi,
a SkL phase appears adjacent to helical and conical phases
\cite{Buhrandt2013}.
Although Gd$_{3}$Ru$_{4}$Al$_{12}$ and MnSi are quite different spin systems, discussion about the TC phase
is important to compare to SkL phase.
The Helix-TC phase transition in Gd$_{3}$Ru$_{4}$Al$_{12}$ was explained by the coexistence of easy axis and easy 
plane types magnetic anisotropies
\cite{Nakamura2023}.
However, the relation between this transition and MQ interactions is not fully elucidated.
In this subsection, we consider this phase transition.
As we mentioned before, MQ interactions do not work between MQ moments with different chiralities. 
Taking this for granted, we omit symbols of chiralities in this subsection to avoid complication.

\begin{figure}[t]
		\includegraphics[width=8.5cm]{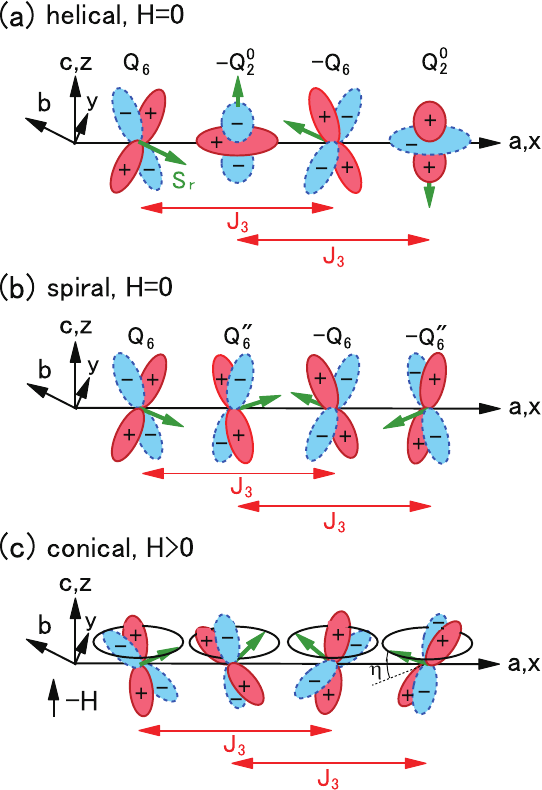}	
		\caption{The structures of ${\bm S_{r}}$ and QSs along the $a$ axis. In this figure,
		we define $x$ and $z$ axes in the directions of $a$, $c$ axes, respectively. The bidirectional red arrows 
		and the sign $J_{3}$ denote AFM dipole interaction.
		(a) The helical structure pulled out from Fig.~\ref{f11}(b). The directions of ${\bm S_{r}}$s are parallel to the 
		$bc$ plane.
		(b) The spiral structure at zero field. The directions of ${\bm S_{r}}$s are parallel to the $ab$ ($xy$) plane.
		(c) The TC structure  which is induced when fields ($-{\bm H} \parallel c$) are
		applied to the structure in Fig.~\ref{f17}(b). The symbol $\eta$ is the angle between the $ab$ plane and ${\bm S_{r}}$s.
		}
		\label{f17}
	\end{figure}

Figure~\ref{f17} displays the structures of ${\bm S_{r}}$s and QSs along the $a$ axis.
As shown in Fig.~\ref{f11}(b), ${\bm S_{r}}$s take the helical structure along the $a$ axis at zero field. 
We redraw it in Fig.~\ref{f17}(a) pulling it out from the figure. In Fig.~\ref{f17}(a), the directions of ${\bm S_{r}}$s
are in the $bc$ plane, and ${\bm S_{r}}$ rotates by $90^\circ$ as it moves trough the site one step to the right.
As described in Eq.~\ref{s4_e53H}, MQ moment
$Q_{6}(\pi/6)$ is the linear combination of $Q_{zx}$ and $Q_{yz}$, therefore, no MQ interactions act between
these moments and $\pm Q_{2}^{0}$s in Fig.~\ref{f17}(a). 
Therefore, quadrupole frustration is eliminated in this structure.
On the other hand, according to Eq.~\ref{s3_e50eHH}, dipole interaction 
$J_{3}$ works between trimers with $\pm Q_{6}$ as illustrated in Fig.~\ref{f17}(a) with bi-directional red arrows. 
The dipole interaction works between trimers with $\pm Q_{2}^{0}$ in this figure also.

Here, we assume a virtual spiral spin structure indicated by green arrows in Fig.~\ref{f17}(b) at zero field, and consider
the Helix-TC transition. We assume this spin structure is associated with QS ordering as shown in this figure.
The directions of ${\bm S_{r}}$s in Fig.~\ref{f17}(b)
are in the $ab$ plane, and ${\bm S_{r}}$ rotates by $90^\circ$ as it moves trough the site one step to the right
along the $a$ axis.
The QSs $Q_{6}$ and $Q_{6}''$ in Fig.~\ref{f17}(b) are respectively defined as
		\begin{align}
			Q_{6}\left(\frac{\pi}{6}\right) &= \left( \cos{\frac{\pi}{6}} \right) Q_{zx} + \left( \sin{\frac{\pi}{6}} \right) Q_{yz},
			 		\label{s4_e71cCc} 
			 	\\
			Q_{6}''\left(\frac{4\pi}{6}\right) &= \left( \cos{\frac{4\pi}{6}} \right) Q_{zx} + \left( \sin{\frac{4\pi}{6}} \right) Q_{yz}
			 		\label{s4_e72dDd}		 
		\end{align}
(Appendix E).
Although not depicted in this figure, both $Q_{6}$s and $Q_{6}''$s are aligned antiferromagnetically
 along the $b$ axis, and ${\bm S_{r}}$s
are aligned ferromagnetically along the $b$ axis. 
MQ interaction between $Q_{6}$ and $Q_{6}''$ is given as
		\begin{align}
			-G_{6}Q_{6}\left(\frac{\pi}{6}\right) Q_{6}''\left(\frac{4\pi}{6}\right) 
				&= -G_{6}\left( \cos{\frac{\pi}{6}} \right)\left(\cos{\frac{4\pi}{6}}\right) ( Q_{zx} )^{2}  \notag \\
				  &\quad -G_{6} \left( \sin{\frac{\pi}{6}} \right) \left( \sin{\frac{4\pi}{6}} \right) (Q_{yz})^{2}  \notag \\
				  &=0.
			 		\label{s4_e73dDd}		 
		\end{align}
On the other hand, each $Q_{6}$ or $Q_{6}''$ in Fig.~\ref{f17}(b) interacts with NN QSs in the  
directions of $b$ axis, respectively.
As we discussed in the previous subsection, ${\bm S_{r}}$s on $Q_{6}$ or $Q_{6}''$ feels easy plane type 
effective anisotropic fields through the MQ interactions. When fields are applied along the $c$ axis, 
the spiral structure in Fig.~\ref{f17}(b) changes into the conical structure shown in Fig.~\ref{f17}(c).

\begin{figure}[t]
		\includegraphics[width=8.5cm]{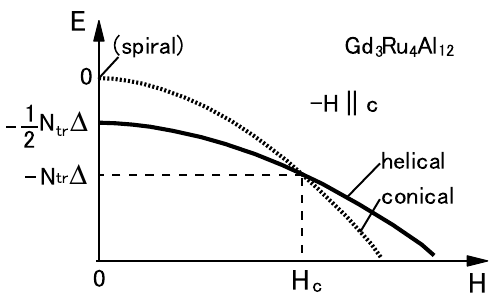}	
		\caption{ Changes in the energy of Gd$_{3}$Ru$_{4}$Al$_{12}$ as functions of applied field ($-{\bm H}\parallel c$) 
		when the field is weak.
		The solid curve and the dotted curve denote the 
		energy of helical structure in Fig.~\ref{f17}(a)  and conical structures in Fig.~\ref{f17}(c), respectively.
		$N_{tr}$ and $\Delta$ denote the number of trimers in a unit volume and easy axis type anisotropic energy,
		respectively. $H_{c}$ denotes the helical-TC transition field. 
		}
		\label{f18}
	\end{figure}

As illustrated in Fig.~\ref{f17}, the numbers of trimers which contribute to the susceptibility in spiral structure is two 
times larger than those in the helical structure when weak applied fields are directed along the $c$ axis. 
Therefore, there is a relation between magnetic susceptibilities of two structures as follows.
		\begin{align}
			\chi_{c}^{spi}(0 \,{\rm K})=2\chi_{c}^{hel}(0 \,{\rm K}). 
			\label{s4_e74gG}	
		\end{align}
Here, the letters on the right shoulders of $\chi$ indicates the helical (hel), spiral (spi) and conical (con) structures.
The changes in the energy of the structures in Fig.~\ref{f17} are illustrated in Fig.~\ref{f18} as functions of applied field 
($-{\bm H}\parallel c$).
In this figure, the zero energy is defined as that of the spiral structure in Fig.~\ref{f17}(b) at zero field. 
The dotted curve in Fig.~\ref{f18} indicates the magnetic field dependence of the energy of the conical structure in 
Fig.~\ref{f17}(c).
In Fig.~\ref{f18}, $N_{tr}$ denotes the number of trimer in a unit volume. 
The energies of Gd$_3$Ru$_4$Al$_{12}$ in the low temperature limit with helical and conical structures in 
fields ($-{\bm H}\parallel c$) are described as
		\begin{align}
			E^{hel}(H) &=-\frac{1}{2}N_{tr} \, \Delta - \frac{1}{2} \mu_{0} \chi_{c}^{hel} H^{2},
			 		\label{s4_e75Hhh} \\
			E^{con}(H) & = -\frac{1}{2}\mu_{0} \chi_{c}^{spi} H^{2}=-\mu_{0} \chi_{c}^{hel} H^{2},
			 		\label{s4_e76IiI}		 
		\end{align}
respectively. Here, $N_{tr}$ is the number of trimer per unit volume, and $\Delta$ is defined as
		\begin{align}
			\Delta&= 2[E^{con}(0) - E^{hel}(0)].
					\label{s4_e77Kkj}
				
		\end{align}
The energy $\Delta$ plays a role as an anisotropic energy in the usual spin flop transitions.
Because the decrease rate in $E^{con}(H)$ is two times greater than that in $E^{hel}(H)$, 
the Helix-TC transition is expected at transition field $H_{c}$ as presented in Fig.~\ref{f18}.
Actually this jump in the magnetization is observed 
at  $\mu_{0}H_{c}=1.25$ T as shown in Fig.~\ref{f13}. 
Referring to Eqs.~\ref{s4_e75Hhh} and \ref{s4_e76IiI},
		\begin{align}
			\Delta& =\frac{1}{N_{tr}}\mu_{0}\chi_{c}^{hel}H_{c}^{2}.		\label{s4_e78Kk} 		
		\end{align}
Substituting $-\mu_{0}H_{c}=1.25$ T (see Fig.~\ref{f13}), 
		\begin{align}
			&\Delta=5.57\times 10^{-23}	\label{s4_e79LLL}
		\end{align}			
in the unit of J, or
		\begin{align}			
			&k_{\rm B}^{-1}\Delta=4.04	\label{s4_e80Mm}.
		\end{align}

	\begin{figure}[b]
		\includegraphics[width=8cm]{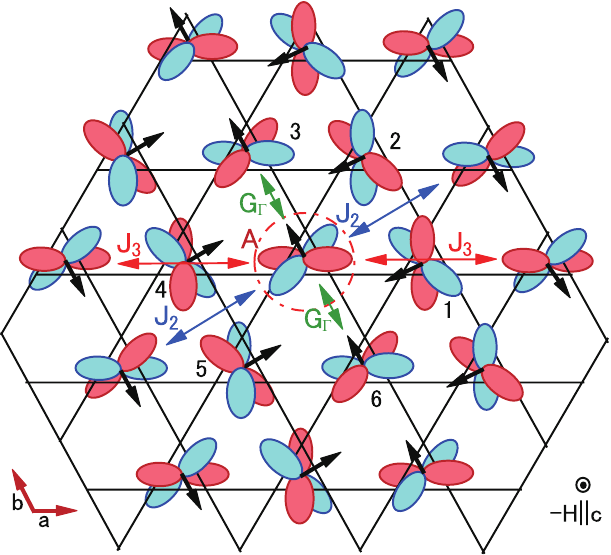}	
		\caption{Structure of ${\bm S_{r}}$s (the bold black arrows) and QSs in TC phase (see Fig.~\ref{f12}). 
		The field is directed along the $c$ axis.
		The angles formed by the $ab$ plane and ${\bm S_{r}}$s [the angle $\eta$ in Fig.~\ref{f17}(c)] are commonly 
		$\pi/6$ (see text).
		The thin blue and red bi-directional arrows indicate the AFM dipole interaction between NNN and 
		3dNN ${\bm S_{r}}$s, respectively.
		The solid green bi-directional arrows denotes AFM MQ interactions. In this figure, only interactions which acts 
		on A site are illustrated. Interactions which are canceled are omitted.
		}
		\label{f19}
	\end{figure}


As shown in Fig.~\ref{f12}, TC phase appears in the intermediate fields directed along the $c$ axis.
In Fig.~\ref{f19}, the structure of ${\bm S_{r}}$s ($S_{r}=15/2$) in the TC phase is illustrated. 
Overlapping this spin structure, we present the QS structure we propose in this figure.
The bold black arrows in Fig.~\ref{f19} denote ${\bm S_{r}}$s on the trimers, and these ${\bm S_{r}}$s form the angle $\eta$
by the $ab$ plane as illustrated in Fig~\ref{f17}(c). According to Figs.~\ref{f12} and \ref{f13}, the lowest field at which TC phase
appears is approximately 1.25 T, and the magnetization at this field (2.5 $\mu_{\rm B}$ per Gd ion) is approximately a half of 
the saturation magnetization expected from the spin system of ``$S=5/2$''.
Therefore, we assume $\eta=\pi/6$. 
Here, we estimate the energy of TC phase at $T=0$.
We focus on trimer ``A'' in Fig.~\ref{f19} surrounded by the dotted-broken red circle.
The thin blue and red bi-directional arrows in this figure are indications of dipole interactions between trimer A and other trimers.
The symbol $J_{2}$ denote dipole interactions between NNN trimers and $J_{3}$ indicate interactions between 3dNN trimers,
and symbol $G_{\Gamma}$ denote MQ interactions. The MQ interactions work between trimer A and surrounding six trimers.
However, the MQ interactions from four trimers (No. 1, 2, 4 and 5 in Fig.~\ref{f19}) among the six trimers are canceled 
with each other, 
and only MQ interactions from two trimers (No. 3 and 6) remain.

Concerning the dipole interactions work on trimer A site, the situation differs depending on the components of dipole interactions.
The components of dipole interactions parallel to the $ab$ plane partly cancel with each other. 
As a result, the components of dipole interactions parallel to the $ab$ plane from two of NNN trimers and from 
two of 3dNN trimers respectively remain.
Only these remained dipole interactions parallel to the $ab$ plane are indicated in Fig.~\ref{f19} with bi-directional blue arrows ($J_{2}$) and with red arrows ($J_{3}$).
On the other hand, the components of dipole interactions parallel to the $c$ axis from six NNN trimers 
stregthen with each other.

Consequently, the change in the energy originating from multipole interactions concerning trimer A in the 
TC phase is given as
		\begin{align}

			E_{TC}^{mp}=&2G_{5} \left(\frac{Q_{5}}{2} \right)^{2} + 2G_{6} \left(\frac{\sqrt{3}Q_{6}}{2} \right)^{2} \notag\\
			&+ 2 (J_{2} + J_{3}) \left( \frac{\sqrt{3}S_{r} }{2}\right)^{2}-6(J_{2}+J_{3})\left(\frac{S_{r}}{2}\right)^{2}. 
			 \label{s4_e81bBbb} 
		\end{align}
The third term on the right side of this equation is the contribution from the component spin parallel to the $ab$ plane, 
and the fourth term on the right side is the contribution from the component spin parallel to the $c$ axis.
These terms are canceled with each other.
Considering duplication, the change in the energy per trimer in TC phase at 1.25 T is obtained as
		\begin{align}
			&E_{TC}^{tr}(0{\rm K},\mu_{0}H_{c}) \notag\\	
			&=\frac{1}{4} G_{5} (Q_{5})^{2} + \frac{3}{4} G_{6} (Q_{6})^{2}  -\frac{1}{2}g_{s}\mu_{\rm B}S_{r}(\mu_{0}H_{c}),
			\label{s4_e82cCCC}		 
		\end{align}
where
the third term on the right side represents the Zeeman energy.	
Because trimers in TC phase are energetically equivalent with each other as shown in Fig.~\ref{f19}, Eq.~\ref{s4_e82cCCC}
represents the change in energy originating from the multiple interactions per trimer at $\mu_{0}H_{c}=1.25$ T. 
We will compare this change in energy to that of SkL in the following section.

\section{The relationship between the Skyrmion lattice and MQ interactions} 

	\subsection{Trimer hexagons as thermally excited local states in fields} 

\begin{figure}[t]
		\includegraphics[width=7cm]{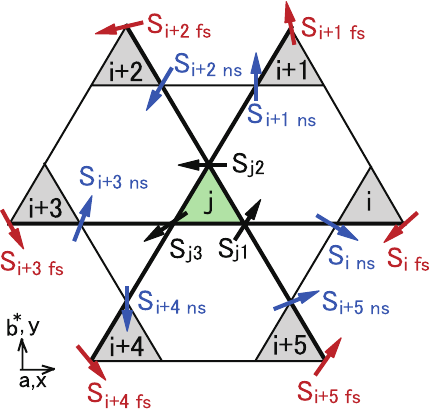}	
		\caption{The placement of the spins in Eq.~\ref{s4_e83dDDd}. 
		The symbols ``$j$'' and ``$i+m$'' ($m=0-5$) represent names of the trimer sites.
		The subscription ``$fs$'' of ${\bm S_{i}}$ denotes the spin which 
		is placed on the far side from $j$-site trimer. The subscription ``$ns$'' of ${\bm S_{i}}$ indicates the spin which 
		is placed on the near side from $j$-site trimer. The bold black lines denote sides of the lattice which are sheared
		by $j$-site and NN-sites triangles. Only blue and red spins on these bold lines contribute to generate 
		spin polarization on $j$-site.
		}
		\label{f20}
	\end{figure}

In the previous section, we discuss the consistency between the Hamiltonian and 
macroscopic magnetic features of Gd$_{3}$Ru$_{4}$Al$_{12}$ previously reported. 
However, from a different perspective, Gd$_{3}$Ru$_{4}$Al$_{12}$ is a host material of skyrmions.
In this section, we discuss the consistency between the SkL and the Hamiltonian.
As we mentioned before, the Hamiltonian (Eq.~\ref{s3_e50eHH}) which is described with ${\bm S_{r}}$ and
 ${\bm Q_{\Gamma\gamma}}$
is effective in FM or AFM quadrupole phases.
According to Eqs.~\ref{eD4}--\ref{eD6}, however, 
the interactions between the trimers may not be described only by ${\bm Q_{\Gamma\gamma}}$ and ${\bm S_{r}}$
when Gd$_{3}$Ru$_{4}$Al$_{12}$ is not in FM or AFM quadrupole phases.
In these cases, we have to take dipole interactions between component spins of trimers into consideration. 
The Hamiltonian is described as
		\begin{align}
			\mathscr{H}_{1}=\mathscr{H}_{0}-J_{1}'\sum_{NN(i,j)}\sum_{k=1}^{3} 
						\left[({\bm S_{i, fs}^{\alpha}}-{\bm S_{i,ns}^{\alpha}})\cdot{\bm S_{jk}}\right],
										\label{s4_e83dDDd}
		\end{align}
where $\mathscr{H}_{0}$ is the Hamiltonian defined in Eq.~\ref{s3_e50eHH}. 
The second term on the right side of this equation is the term which cannot be described using 
resultant spin ${\bm S_{r}}(i)$.
The symbol $J'_{1}$ in Eq.~\ref{s4_e83dDDd} denotes the coupling constant, and 
the sign is expected to be positive (Appendix D). 
The spacial arrangement of the component spins ${\bm S_{i, fs}^{\alpha}}$,
${\bm S_{i, ns}^{\alpha}}$ and ${\bm S_{jk}}$ in this equation are illustrated in Fig.~\ref{f20}. 
The superscripts $\alpha$ of these spins indicate the planes perpendicular to ${\bm S_{r}}(i)$. The subscriptions
``$ fs$'' and ``$ ns$'' of ${\bm S_{i}}^{\alpha}$ denote the far side and near side from the $j$-site, respectively 
(see Fig.~\ref{f20}).
The symbols ${\bm S_{jk}}$ ($k=1,2,3$) in Eq.~\ref{s4_e83dDDd} denote the component spins on the $j$-site trimer
in Fig.~\ref{f20}. Only the spins represented in Fig.~\ref{f20} are included in the Hamiltonian $\mathscr{H}_{1}$.

\begin{figure}[t]
		\includegraphics[width=7.5cm]{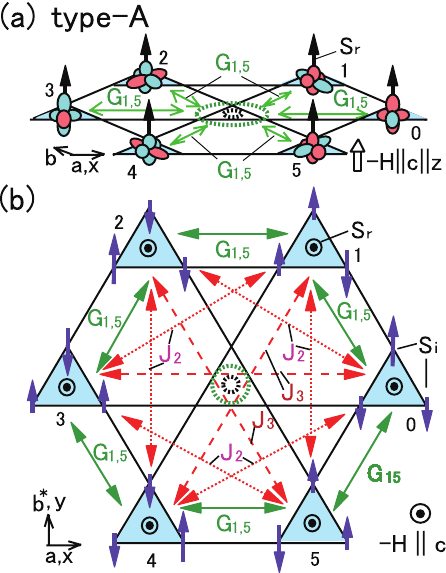}	
		\caption{The type-A TrH and a ST under fields directed along the $c$ axis.
		The numbers of the trimers $n_{t}$ (0 to 5) are described in this figure.
		(a) A view of angle perspective. The bunches of grapes shaped
		pictures denote QSs. The dotted green and dotted black circles on the ST
		are indications of quadrupole disorder and dipole disorder, respectively. (b) A project drawing along the direction
		of $c$ axis. The symbols $G_{1,5}$ denote the coexistence of MQ interactions $G_{1}$ and $G_{5}$.
		The symbols $J_{2}$ and $J_{3}$ indicate dipole interactions in Eq.~\ref{s3_e50eHH}. 
		The symbol $\odot$ indicate directions of ${\bm S_{r}}$s.
		The bold purple arrows are indications of component spins of $ab$ plane ${\bm S_{i}^{xy}}$s.
		The lengths of short bold purple arrows are $1/2$ of those of long bold purple arrows.
		The dipole interactions between NN 
		described in Eq.~\ref{s4_e83dDDd} are canceled with each other on each trimer.
		}
		\label{f21}
	\end{figure}

First, we consider local spin structures illustrated in Fig.~\ref{f21} as a
thermally excited state.  We limit the discussion only to the cases in which applied fields
are directed along the $c$ axis.
Figure~\ref{f21}(a) shows the trimer hexagon (TrH) of type-A from an angle perspective. 
We put numbers $n_{t}$ from 0 to 5 on the trimers in this figure.
The bi-directional green arrows labeled ``$G_{1,5}$'' in Fig.~\ref{f21} indicate AFM MQ interactions with $\Gamma_{1}$
and $\Gamma_{5}$ symmetry as we mentioned in the previous section. The bold black arrows in this figure denote ${\bm S_{r}}$s.
Figure~\ref{f21}(b) displays a project drawing of  Fig.~\ref{f21}(a)
along the $c$ axis. The dotted bi-directional red arrows labeled ``$J_{2}$'' and the broken bi-directional red arrows
labeled ``$J_{3}$'' in this figure indicate dipole interactions between NNNs and 3dNNs which are described in 
Eq.~\ref{s3_e50eHH}, respectively.

According to the easy axis anisotropy, the directions of ${\bm S_{r}}$s are parallel to
the $c$ axis. The QSs of the trimers in Fig.~\ref{f21} are described as
		\begin{align}
			Q_{1,5}(n_{t})&
				= (-1)^{n_{t}} \bigg[ \left(\frac{Q_{2}^{0}+Q_{2}^{2}}{2}\right) \cos{\frac{2n_{t}\pi}{3}} \notag \\
						&\qquad\qquad\quad - \left(\frac{R_{m}+Q_{xy}}{2}\right) \sin{\frac{2n_{t}\pi}{3}} \bigg] 
			 		\label{s5_e84eEE}		 
		\end{align}
(Appendix E).
The dotted green and the dotted black circles on the single trimer (ST) site placed at the center of
the TrH in Fig.~\ref{f21}(a) denote disorder of QS and disorder of ${\bm S_{r}}$, respectively. 
The ST site does not belong to TrH,
and it becomes the quadrupole disordered site because 
the MQ interactions from the surrounding six QSs are canceled on this site. 
Similar to the MQ interactions, dipole interactions from the surrounding six dipole moments are also canceled 
on the ST site, and this site becomes a dipole disordered site also.

The bold purple arrows in Fig.~\ref{f21}(b) indicate component spins on the trimers.
Here, we focus on the trimer site No. 1 in Fig.~\ref{f21}(b). 
The dipole interactions from NN (sites No. 0 and No. 2) in both sides described by the second term in the right side 
of Eq.~\ref{s4_e83dDDd} 
are canceled with each other on site No. 1. 
This situation is similar on other trimer sites in the TrH.
Consequently, the energy of type-A TrH is given as
		\begin{align}
			E_{A}=& -\frac{3}{4}G_{1}(Q_{1})^{2} -\frac{3}{4}G_{5}(Q_{5})^{2} 
			- 6J_{2}S_{r}^{2} -3J_{3}S_{r}^{2} \notag \\ 
			&-6g_{s}\mu_{\rm B}S_{r}(\mu_{0}H),
			 		\label{s5_e85dDFf}		 
		\end{align}
where $S_{r}=15/2$.
The first and second terms in the right side of this equation is originated from MQ interaction between the NN trimers,
and the third term and the fourth terms are originated from dipole interactions between NNN trimers and 3dNN trimers, 
respectively. The fifth term in the right side of Eq.~\ref{s5_e85dDFf} is the Zeeman energy originated from coupling of
${\bm S_{r}}$s and applied magnetic fields. As described in Eq.~\ref{s5_e85dDFf}, the TrH is an eigenstate of
the Hamiltonian given in Eq.~\ref{s3_e50eHH} as a result. The formation of TrH causes energy loss arising from 
AFM MQ interactions and dipole interactions.
On the other hand, the formation of TrH leads to energy gain arising from Zeemann energy.
At the same time, TrHs keep disorder on STs placed in them.
Because the degrees of freedom of QS is large, this shielding of ST make a major contribution to the decrease
in free energy.


%
	\begin{figure}[t]
		\includegraphics[width=7.5cm]{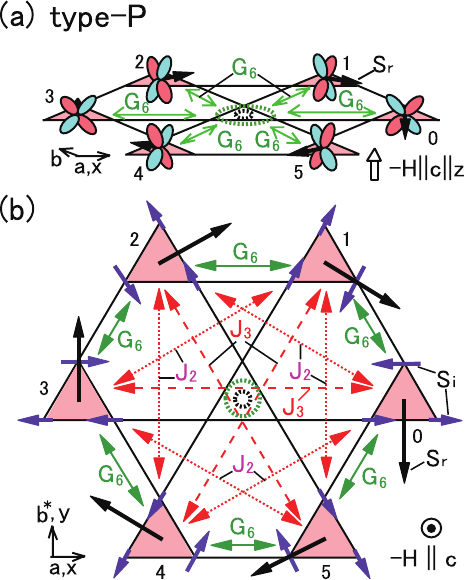}	
		\caption{The type-P TrH and a ST under fields directed along the $c$ axis.
		The bold black arrows denote ${\bm S_{r}}$s and the butterfly shaped
		pictures denote QSs.
		The numbers of the trimers $n_{t}$ (0 to 5) are described in this figure.
		(a) A view of angle perspective. The bold black arrows indicate ${\bm S_{r}}$s,  The dotted green and dotted 
		black circles on the ST are indications of quadrupole and dipole disorder, respectively. 
		(b) A project drawing along the direction of $c$ axis.
		The purple arrows denote component spins ${\bm S_{i}}$s which is parallel to the $ab$ plane.
		The bidirectional dotted red and broken red arrows indicate dipole interactions between NNN and 3dNN
		described in Eq.~\ref{s3_e50eHH}, 
		respectively. The spin polarizations originating from dipole interactions between NN 
		described in Eq.~\ref{s4_e83dDDd} are parallel to the ${\bm S_{r}}$s.
		}
		\label{f22}
	\end{figure}

We consider another type (P-type) of TrH illustrated in Fig.~\ref{f22} which appears in the fields directed 
along the $c$ axis. We put numbers $n_{t}=0-5$ by the trimers in this figure.
As shown in Fig.~\ref{f22}(a), ${\bm S_{r}}$s are directed in the $ab$ plane.
The dotted green and black circles on the ST site placed on the center of the TrH denote quadrupole and dipole 
disorder, respectively. This ST is not the component of the TrH.
Similar to type-A TrH, MQ interactions from the TrH in Fig.~\ref{f22}(a) are canceled on the ST site. 
The QSs on the number $n_{t}$ trimer in this figure are described as
		\begin{align}
			Q_{6}(n_{t})&
				=R_{z}\left(\frac{n_{t}\pi}{3}\right)\left(\frac{Q_{zx}^{+}+Q_{zx}^{-}}{2}\right),
			 		\label{s5_e86eEEf}		 
		\end{align}
where $R_{z}(n_{t}\pi/3)$ denotes rotation around the $z$ axis by angle $(n_{t}\pi)/3$. 
Figure~\ref{f22}(b) shows project drawing of type-P TrH along the $c$ axis. 
Each trimer on the TrH receives MQ interactions from both sides of NN trimers.
The bold purple arrows which are parallel to the $ab$ plane denote component spins ${\bm S_{i}}$s of the trimers.
The dotted red and broken red bi-deirectional arrows in
Fig.~\ref{f22}(b) are indications of dipole interactions between NNN and 3dNN trimers, respectively. 
The symbols $J_{2}$ and $J_{3}$ in this figure denote coupling constants in Eq.~\ref{s3_e50eHH}.

In the case of type-P TrH, the dipole interactions between NN trimer sites deduced from the second term 
on the right side of Eq.~\ref{s4_e83dDDd} cannot be ignored. Here, we focus on No. 1 trimer in Fig.~\ref{f22}(b).
This trimer receives dipole interactions from No. 0 and No. 2 trimers. 
As shown in Fig.~\ref{f22}(b),
these interactions are not canceled on No. 1 trimer. However, the synthesized spin polarization on No. 1 trimer
induced by No. 0 and No. 2 trimers is parallel to the direction of ${\bm S_{r}}$ on No. 1 trimer (Appendix F).
On the other hand, the synthesized effective dipole field on the No. 1 trimer originated from the dipole interactions 
from No. 2 trimer and No. 0 trimers is anti-parallel to the direction of ${\bm S_{r}}$ on No.1 trimer.
Thus, the energy of type-P TrH is given as
		\begin{align}
			E_{P}&= -6 G_{6} \left[\frac{(Q_{zx}^{+})^{2}+(Q_{zx}^{-})^{2}}{4}\right]\cos{\frac{\pi}{3}}
			 \notag\\
			&\qquad- 6J_{2}S_{r}^{2}\left(\cos{\frac{2\pi}{3}}\right)+3J_{3}S_{r}^{2}
				\label{s5_e87FffF} \\
			&= -\frac{3}{2}G_{6}(Q_{6})^{2} + 3J_{2}S_{r}^{2} + 3J_{3}S_{r}^{2},		 		
					\label{s5_e88FGGg}
		\end{align}
where $S_{i}=7/2$ and $S_{r}=15/2$.
Here, we ignore Zeeman energy of ${\bm S_{r}}$s as an approximation because of easy plane anisotropy in the trimers. 
It can be seen from Eq.~\ref{s5_e88FGGg}, the formation of type-P TrH causes energy loss by the MQ interaction
and causes energy gain by the dipole interactions.

	\subsection{Lattice formation of TrHs  and relation to SkL}

\begin{figure*}[t]
		\includegraphics[width=17cm]{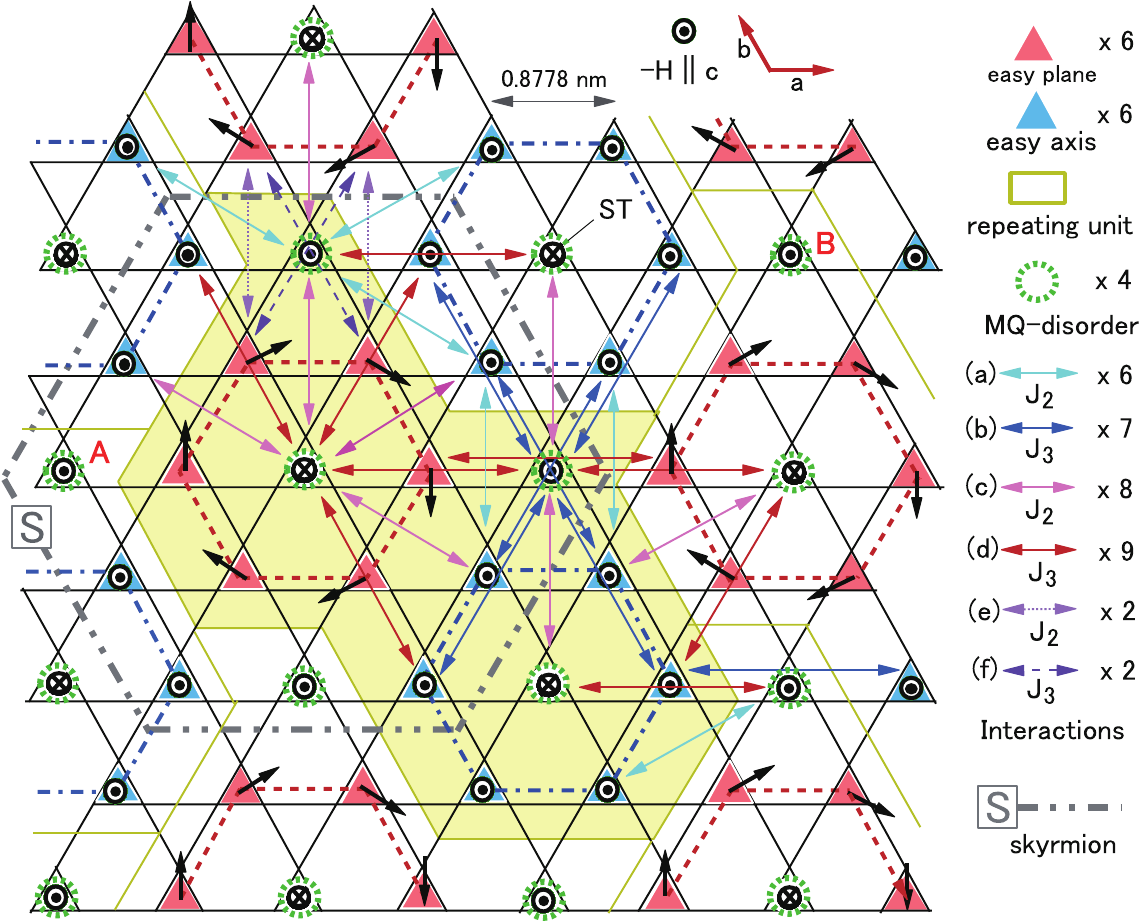}	
		\caption{The lattice formed of TrHs. The dotted broken blue hexagon and broken red hexagon indicate
		TrHs of type-A and type-P, respectively.  The symbols ``ST'' denote single trimers which are not included 
		in TrHs (see text). This figure contains sventeen STs totally. 
		The solid bold black arrows, black $\odot$ and $\otimes$ denote 
		${\bm S_{r}}$s, and these symbols represent the directions of ${\bm S_{r}}$s. 
		The symbols $J_{2}$, $J_{3}$ and $G_{6}$ denote coupling constants in Eq.~\ref{s3_e50eHH}. 
		The area painted with light yellow is the indication of repeating unit of the TrH lattice.
		The dotted green circles denote MQ disorder sites. The thin bidirectional arrows indicate dipole
		interactions in Eq.~\ref{s3_e50eHH}. The difference in colors of these arrows corresponds to the difference of
		energies originating from the difference in distance or in relative directions of two ${\bm S_{r}}$s connecting with arrows
		(see Table~\ref{Tab3}). 
		The double dotted broken green lines with symbol ``S'' surrounded by the square denotes a skyrmion particle.
		For the sake of easy look, only one skyrmion particle is indicated in this figure.
		Concerning the quadrupole disorder on ``A'' and ``B'' sites, see Appendix G.
		}
		\label{f23}
	\end{figure*}

In this subsection, we consider a lattice formed by the type A and type P TrHs as illustrated in Fig.~\ref{f23}. 
This lattice enfolds STs in it.
In Fig.~\ref{f23}, type-A TrHs are indicated by the dotted broken blue hexagons, and
type-P TrHs are represented with broken red hexagons.
The thin blue and thin red triangles painted denote trimers which show easy axis anisotropy and easy plane anisotropy,
respectively. 
As shown in Fig.~\ref{f23}, the closest distances between two TrHs which belong to the same type are larger than the
distance between NN trimers. Therefore, these TrHs are not interact in the quadrupole manner.
The closest distances between two TrHs which belong to different types are within range of NN, however, no quadrupole
interactions work between these TrHs because of the difference in symmetry.
Thus, no quadrupole interactions work between the TrHs in Fig.~\ref{f23}.
This keeps the stability of each TrH in the lattice, and they look as particles in the lattice.
The dotted green rings in Fig.~\ref{f23} denote quadrupole disorder on the trimer sites. As can be seen from Figs.~\ref{f21} and 
\ref{f22}, quadrupole interactions are canceled on the ST site which placed on the center of the TrH. 
The solid yellow lines in Fig.~\ref{f23} represent the boundary of a repeating unit of the lattice, and
this area is painted with thin yellow. Each repeating unit contains a pair of TrHs of two types and 
four STs.

When the TrHs form the lattice and enfolds STs, the energy of the lattice is expected to change from the total amount 
of the energiesof isolated TrHs and STs due to the dipole interactions among them. 
Here, we examine this change in energy.
The thin bi-directional arrows in Fig.~\ref{f23}
indicate pairs of ${\bm S_{r}}$s which interact with each other through dipole interactions. 
In this figure, for the purpose of calculating the energy
per single repeating unit painted thin yellow, we only display dipole interactions not duplicated. 
In Fig.~\ref{f23}, there are six kinds of non-equivalent pairs labeled ``a--f'' distinguished by the kinds of lines and colors 
of the bi-directional arrows. 
The change in energy of these pairs are the same when the bi-directional arrows 
show the same kind.
We summarize the kinds of pairs, change in the energy by the dipole interactions and quantity of 
pairs (quantity of the bi-directional arrows) in Table~\ref{Tab3}. According to this table, the change in the energy by 
forming the lattice is obtained as
		\begin{align}
			E_{L}^{ru}&=3\left(J_{2}+J_{3}\right)S_{r}^{2} 
			 		\label{s3_e89GGg}		 
		\end{align}
per the repeating unit area. 
Because both signs of $J_{2}$ and $J_{3}$ are negative, the system gets energy gain with forming the lattice, therefore,
this energy can be regarded as a ``cohesive energy'' of TrHs per repeating area.
The total Zeeman energy of STs in the repeating area is zero
as we can see from Fig.~\ref{f23}.
As indicated by dotted green circles in Fig.~\ref{f23}, STs still remain quadrupole 
disorder because of cancellation of the quadrupole interactions (Appendix G).

		\begin{table}[t]
			\caption{Dipole interactions between ${\bm S_{r}}$s represented by bi-directional arrows
			 in Fig.~\ref{f23}. This table includes 
			change in energy by these dipole interactions and quantities of pairs equivalent in the repeating unit
			painted with thin yellow.}							
			\center					
			\begin{tabular}{ccc}
			\hline 
			interactions 	& \quad change in energy 	& \quad quantities of pairs	\\
			\hline 	
			a	& \quad	$-J_{2} S_{r}^{2}$		& \quad	6 \\
			b	& \quad	$-J_{3} S_{r}^{2}$		& \quad	7 \\
			c	& \quad	$J_{2} S_{r}^{2}$		& \quad	8 \\
			d	& \quad	$J_{3} S_{r}^{2}	$		& \quad	8 \\
			e	& \quad	$(1/2)J_{2} S_{r}^{2}$	& \quad	2 \\
			f	& \quad	$J_{3} S_{r}^{2}$		& \quad	2 \\	
			\hline
			\end{tabular}
			\label{Tab3}
		\end{table}

Here, we discuss the relationship between the TrH lattice and SkL.
If we change our perspective, the lattice of TrHs enfolds STs in Fig.~\ref{f23} looks as a SkL. 
The double dotted broken gray lines with
the symbol ``S'' surrounded by the square in this figure shows some features of a skyrmion particle.
On the center of this particle, the ${\bm S_{r}}$ is directed to the back to the paper.
As it moves from the center, the directions of ${\bm S_{r}}$s become parallel to the paper, and
${\bm S_{r}}$s are directed to the front of the paper on the periphery of the particle.
Therefore, the particle can be regarded as a skyrmion particle.
For the sake of easy look, we do not indicate all skyrmion particles in Fig.~\ref{f23} with broken gray lines, 
but these particles are formed
around  P-type TrHs, and form the SkL in the $ab$ plane.
As illustrated in Fig.~\ref{f23}, the diameter of the skyrmion in this figure is estimated to be 4.4 nm approximately, and
the circular shape structures of ${\bm S_{r}}$s around the center of skyrmion particles
rotate in the same direction.
These features approximately agree with the spin structure of SkL reported in the previous study
\cite{Hirschberger2019_2}.
Electrostatic energies play important roles to form the circular shape of ${\bm S_{r}}$s in the $ab$ plane (Appendix F).
The same direction in these circular shapes
implies that the clockwise and anti-clockwise domains are generated in actual SkL phase.

Lastly, we discuss the thermodynamic stability of SkL phase. 
To compare to the TC phase, we set magnetic fields as $\mu_{0}H_{c}=1.25$ T (Eq.~\ref{s4_e82cCCC}) commonly in these
two phases.. 
According to Eqs.~\ref{s5_e85dDFf}, \ref{s5_e88FGGg} and \ref{s3_e89GGg}, the energy gain due to the
interactions across the repeating unit area in Fig.~\ref{f23} is given as
		\begin{align}
			&E_{SkL}^{ru}=E_{ A} + E_{P} + E_{L}^{ru}  \notag \\
			&=-\frac{3}{4}G_{1}(Q_{2}^{0})^{2}-\frac{3}{4}G_{5}(Q_{xy})^{2}  - 6J_{2}S_{r}^{2} -3J_{3}S_{r}^{2} 
				\notag \\ &\quad -6g_{s}\mu_{\rm B}S_{r}(\mu_{0}H_{c}) \notag \\
			&\quad - \frac{3}{2}G_{6}(Q_{zx})^{2} + 3J_{2}S_{r}^{2} + 3J_{3}S_{r}^{2} \notag\\
			&\quad +3\left(J_{2}+J_{3}\right)S_{r}^{2} 
						 		\label{s5_e90AAa} \\
			&=-\frac{3}{4}G_{1}(Q_{2}^{0})^{2}-\frac{3}{4}G_{5}(Q_{5})^{2} -\frac{3}{2}G_{6}(Q_{6})^{2} + 3J_{3}S_{r}^{2}  \notag \\
			& \quad -6g_{s}\mu_{\rm B}S_{r}(\mu_{0}H_{c}).
						 		\label{s5_e91BbBb}		 
		\end{align}
The repeating unit area contains 16 trimers as shown in Fig.~\ref{f23}. Therefore, energy per trimer in SkL phase is obtained as
		\begin{align}
			E_{SkL}^{tr}&=-\frac{3}{64}G_{1} (Q_{2}^{0})^{2} - \frac{3}{64}G_{5}(Q_{5})^{2}
						+\frac{3}{32}G_{6} (Q_{6})^{2} \notag \\
						&\quad + \frac{3}{16} J_{3}S_{r}^{2} 
							-\frac{3}{8} g_{s}\mu_{\rm B}S_{r}(\mu_{0}H_{c}). 
						 		\label{s5_e92CCcc}
		\end{align}
Referring to Eqs.~\ref{s4_e82cCCC} and \ref{s5_e92CCcc}, the difference in energy per trimer between SkL phase and TC phase 
under the same field $H_{c}$ (Fig.~\ref{f19}) is given as
		\begin{align}
			&\Delta E^{tr}(\mu_{0}H_{c}) \equiv E_{SkL}^{tr}(\mu_{0}H_{c})-E_{TC}^{tr}(\mu_{0}H_{c})  
								\label{s5_e93DDdd} \\
			=&-\frac{3}{64}G_{1}(Q_{2}^{0})^{2}-\frac{19}{64}G_{5}(Q_{5})^{2} -\frac{21}{32}G_{6}(Q_{6})^{2} \notag \\
			& \quad+ \frac{3}{16}J_{3}S_{r}^{2} 
					+\frac{1}{8}g_{s}\mu_{\rm B}S_{r}(\mu_{0}H_{c}).
						 		\label{s5_e94EEee} \\
			=&\left(-\frac{3}{16}G_{1}-\frac{19}{16}G_{5}-\frac{21}{32}G_{6}\right) (Q_{6})^{2}  \notag\\
			& \quad + \frac{3}{16}J_{3}S_{r}^{2}
			+\frac{1}{8}g_{s}\mu_{\rm B}S_{r}(\mu_{0}H_{c}).
									 	\label{s5_e95EFf} 
		\end{align}
It can be seen from this formula that the SkL loses energy due to MQ interactions but gains energy
due to dipole interactions compare to TC phase. The spin structure of SkL is advantageous to get
energy gain with AFM dipole interactions with finite reaching distances.

It has been pointed out that the degrees of quadrupole freedom of the trimers possibly remains even in dipole ordered state
in Gd$_{3}$Ru$_{4}$Al$_{12}$
\cite{Nakamura2023}.
However, specific effects of that have not been discussed.
As illustrated in Fig.~\ref{f23}, there are 4 quadrupole disordered STs in the unit area.
On these sites, the directions of ${\bm S_{r}}$s are fixed to the direction of $c$ axis, but STs still remain degree of 
quadrupole freedom.
When ${\bm S_{r}}$s are directed along the $c$ axis,
possible QS or $R_{m}$ on these trimer sites are $Q_{2}^{0}$, $R_{m}$, $Q_{2}^{2}$ and $Q_{xy}$, 
and each QS or $R_{m}$ possesses 2 degrees of freedom. 
Therefore, 8 degrees of quadrupole or rotational freedom are remained per one ST totally.
Thus, the free energy of SkL phase per trimer is given as
		\begin{align}
			F_{SkL}^{tr}(T,\mu_{0}H_{c})&=  E_{SkL}^{tr}(\mu_{0}H_{c})-\frac{1}{4}k_{\rm B}T\,\ln{8}. 
						 		\label{s5_e96FGff}
		\end{align}
Strictly speaking, information of collective excitations is needed to calculate free energy of TC phase accurately.
However, contribution to the free energy arising from collective excitations would be small at fully lower 
temperatures than $T_{1}$ or $T_{2}$.
Therefore, we assume the free energy of TC phase is temperature independent here. Thus,
		\begin{align}
			F_{TC}^{tr}(T,\mu_{0}H_{c})&\simeq E_{TC}^{tr}(\mu_{0}H_{c}). 
						 		\label{s5_e97GgGg}
		\end{align}
Then,
		\begin{align}
			&\Delta F^{tr} \equiv F_{SkL}^{tr}(T,\mu_{0}H_{c})-F_{TC}^{tr}(T,\mu_{0}H_{c}) \notag \\
			&\simeq \Delta E^{tr}(\mu_{0}H_{c})-\frac{1}{4}k_{\rm B}T\,\ln{8}.
						 													\label{s5_e98HhHh}
		\end{align}
We try to find specific values of $\Delta E^{tr}$ and $F_{SkL}^{tr}$ at field 1.25 T.
Substituting $\mu_{0}H_{c}=1.25$ T, $Q_{6}=-2.121$, $S_{r}=15/2$, $S_{i}=7/2$ and $\xi=0.6999$ into Eq.~\ref{s5_e95EFf},
		\begin{align}
			&k_{\rm B}^{-1}\Delta E^{tr}(1.25\,{\rm T})	\notag\\
			&=k_{\rm B}^{-1} \left[ -0.8435G_{1} - 5.343G_{5} -2.952G_{6} +10.55J_{3}\right]\notag\\
			&\quad  +1.574. 
						 		\label{s5_e99FGgG}
		\end{align}
Here, we assume the set of coupling constants as
		\begin{align}
				&\begin{pmatrix}
				k_{\rm B}^{-1}G_{1},&k_{\rm B}^{-1}G_{5},	& k_{\rm B}^{-1} G_{6},  &k_{\rm B}^{-1}J_{3}	
				\end{pmatrix}		\notag\\
				&=
				\begin{pmatrix}
					-1.3,		&  	-0.25,		&	-1.45,		&	-0.55			
				\end{pmatrix}.
				\label{s5_e100HHHhh}
		\end{align}
for example. These coupling constants satisfy the conditions in Eq.~\ref{s4_e69gGg}.

\begin{figure}[t]
		\includegraphics[width=8.5cm]{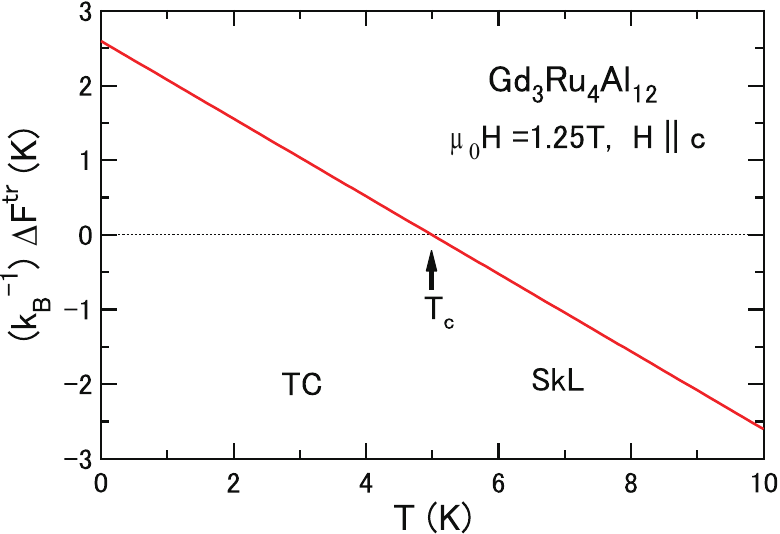}
		\caption{The temperature dependence of the difference in the free energies $k_{\rm B}^{-1}\Delta F^{tr}$ 
		calculated according to Eqs.~\ref{s5_e98HhHh} and \ref{s5_e99FGgG} 
		with the parameters described in Eq.~\ref{s5_e100HHHhh}.
		The magnetic field $\mu_{0} H=1.25$ T is applied along the $c$ axis. The symbol $T_{c}$ (=5.0 K) indicates
		the transition temperature between TC and SkL phases.
		}
		\label{f24}
	\end{figure}

According to Eqs.~\ref{s5_e98HhHh} and \ref{s5_e99FGgG}, we calculate $\Delta F^{tr}$ at 1.25 T as a function of temperature 
using coupling constants in Eq.~\ref{s5_e100HHHhh}.
Figure~\ref{f24} shows the temperature dependence of $\Delta F^{tr}$ in Eq.~\ref{s5_e98HhHh} calculated.
As shown in this figure, $\Delta F^{tr}$ decreases with increasing temperature and changes sign from positive to 
negative at $T_{c}=5.0$ K. 
This insists that the phase transition from TC to SkL occurs at $T_{c}=5.0$ K.
As shown in Fig.~\ref{f12}, this $T_{c}$ calculated approximately agrees with experimental results.
The decreasing rate of $|k_{\rm B}^{-1}\Delta F^{tr}/T|$ in Fig.~\ref{f24} is 
is obtained to be a great value $(1/4)k_{\rm B}\ln{8}=0.52$.
Because the MQ moments are higher rank moments, they possess large amount of degrees of freedom 
when the fields are applied along the $c$ axis. 
Adding this, the structure of the SkL can enfold many disordered STs in it due to the mixed structure
of two types TrHs. 
This large entropy leads to the stability of the SkL phase in intermediate temperatures.
Because the QSs do not couple with magnetic fields directly, the entropy arising from QS
is not  strongly reduced by applied magnetic field. This may be one of the reasons why SkL phase becomes stable in the
intermediate field range.
The appearance of the SkL phase  in Gd$_{3}$Ru$_{4}$Al$_{12}$ closely related to the MQ
moments of trimers and MQ interactions among them.

%


\section{Summary} 

We have determined symmetric structures of imperfect FM trimer spins in Gd$_{3}$Ru$_{4}$Al$_{12}$ which 
possess MQ moments.
We have derived the MQ interactions of the RKKY mechanism in Gd$_{3}$Ru$_{4}$Al$_{12}$, 
and proposed the Hamiltonian which includes both magnetic dipole and MQ interactions. 
One of the characteristics
of this Hamiltonian is ignoring dipole interactions between NN trimers because of the interference
of the RKKY interactions. On the other hand, we assume MQ interactions between NN trimer sites. 
The trimers possesses both magnetic dipole and MQ moments.
Gd$_{3}$Ru$_{4}$Al$_{12}$ exhibits various phase transitions due to degrees of these freedom. 
Our model shows consistency to some features of Gd$_{3}$Ru$_{4}$Al$_{12}$ observed, such as 
successive phase transitions, coexistence of different kinds of magnetic anisotropies, the Helical-TC
phase transition.

Based on the Hamiltonian, we considered excited states in magnetic fields.
We have proposed two types of TrHs as thermally excited states under fields parallel to the $c$ axis.
These TrHs represents easy plane and easy axis magnetic anisotropies, respectively, and
form the lattice of TrHs in the $ab$ plane. 
Once, this lattice of TrHs is formed, it becomes the SkL simultaneously. 
This SkL includes many quadrupole disordered ST sites due to the coexistence of different types of TrHs.
The remaining degrees of quadrupole freedom of these STs reduces the free energy of SkL and stabilize the SkL 
at finite temperatures. 
The MQ degeneracy does not easily lifted by magnetic applied field compare to the magnetic dipole degeneracy
because the MQ moments do not couple with magnetic fields directly. This would be one of the reasons
why the SkL phase appears in the intermediate field range.

In the present study, we have succeeded to explain some properties of Gd$_{3}$Ru$_{4}$Al$_{12}$
assuming Hamiltonian including MQ interactions in classical description. 
The comprehensive and accurate quantitative reproduction
of the properties, and quantized description of the present study are future problems.

\section*{Acknowledgement} 

We are deeply grateful to emeritus professor A. Ochiai (Tohoku University) and Dr. N. Kabeya (Tohoku University) for giving us the opportunity to the present study and for fruitful discussion about frustrated spin systems of rare earth compounds.


\appendix  
\section{} 
The functions $F_{\Gamma \gamma}$s are orthogonal with each other, namely,
		\begin{align}
			\int_{-\infty}^{\infty} F_{\Gamma \gamma}({\bm r}) F'_{\Gamma \gamma}({\bm r}) R(r) d{\bm r}=0
			\label{eA1}
		\end{align}
when $F_{\Gamma \gamma}$ and $F'_{\Gamma \gamma}$ are different. Here, $R(r)$ is a positive number function
of $r$ only, and it rapidly decreases with increasing $r$ when $r$ is fully bigger than the radius of the trimer.
We summarize the point group symmetries of  $F_{\Gamma \gamma}({\bm r})$ in Table~\ref{Tab4}.
	\renewcommand{\arraystretch}{1.5}								
		\begin{table}[h]
			\caption{Point group symmetries of $F_{\Gamma\gamma}({\bm r})$s in hexagonal crystals.}							
			\center					
			\begin{tabular}{ccccc}
			\hline 
			$F_{2}^{0}({\bm r})$	&\quad$F_{2}^{2}({\bm r})$	&\quad$F_{xy}({\bm r})$	
			&\quad$F_{yz}({\bm r})$	&\quad$F_{zx}({\bm r})$ \\
			\hline 
	\renewcommand{\arraystretch}{1} \\			
			$\Gamma_{1}$& \quad$\Gamma_{5}$& \quad$\Gamma_{5}$&	\quad$\Gamma_{6}$& \quad$\Gamma_{6}$ \\		
			\hline
			\end{tabular}
			\label{Tab4}
		\end{table}

\section{} 
\begin{figure}[t]
		\includegraphics[width=5cm]{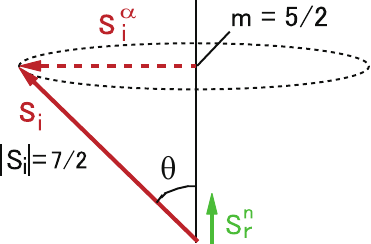}
		\caption{The component spin ${\bm S_{i}}$ and ${\bm S_{i}^{\alpha}}$ under the condition of the minimum angle $\theta$.
				The quantization axis is parallel to ${\bm S_{r}^{n}}$. $m$ is the magnetic quantum number.
		}
		\label{f25}
	\end{figure}

As mentioned in the main text, we treat spins of Gd$^{3+}$ $(S=7/2)$ as classic spins basically, but we need to consider
 quantum nature
of the spins to determine absolute values of MQ moments.
Figure~\ref{f25} shows the component spin ${\bm S_{i}}$ and its $\alpha$ plane component ${\bm S_{i}^{\alpha}}$
under the condition that minimum angle $\theta$ is realized. The quantization axis is directed along the ${\bm S_{r}^{n}}$.
When $\theta$ is minimum, the magnetic quantum number of the ${\bm S_{i}}$ is $m=5/2$. The magnitude of ${\bm S_{i}^{\alpha}}$
is described as
		\begin{align}
			S_{i}^{\alpha}=(7/2)\sin{\theta}=(7/2)\xi.
			\label{eB1}
		\end{align}
Here, $S_{i}^{\alpha}$, $\theta$ and $\xi$ are obtained to be $\sqrt{6}$, 44.42$^\circ$ and $0.6999$, respectively. 
Corresponding magnitude of the magnetic moment is given as
		\begin{align}
			\mu_{i}^{\alpha}&=-g_{s}\mu_{\rm B} S_{i}^{\alpha} 			\label{eB2}\\
					&=-7\mu_{\rm B} \xi.
			\label{eB3}
		\end{align}
Concerning the resultant spin ${\bm S_{r}}$, because ${\bm S_{r}}={\bm S_{1}}+{\bm S_{2}}+{\bm S_{3}}$,
		\begin{align}
			S_{r}=\frac{15}{2}.
			\label{eB4}
		\end{align}

We summarize numerical values of $\tilde{Q}_{\Gamma\gamma}$ in Table~\ref{Tab5}.
We also summarize the eigenvalues of the operators ${\bm Q_{\Gamma\gamma}}$ defined in Eqs.~\ref{s3_e23ff} and
 \ref{s3_e25gg} in Table~\ref{Tab6}.

	\renewcommand{\arraystretch}{1.5}								
		\begin{table}[t]
			\caption{The absolute values of $\tilde{Q}_{\Gamma\gamma}$ and $\tilde{R}_{m}$
			defined in Eqs.~\ref{s2_e12e} and \ref{s2_e20c}
			in the classical spin approximation. The unit in this table is
			$10^{-32}$ ${\rm JT^{-1}m}$.
			}							
			\center					
			\begin{tabular}{cccccc}
			\hline 
			$\tilde{Q}_{yz}^{\pm}$	&\quad$\tilde{Q}_{zx}^{\pm}$	&\quad$\tilde{Q}_{xy}$	
			&\quad$\tilde{Q}_{2}^{2}$	&\quad$\tilde{Q}_{2}^{0}$ &\quad$\tilde{R}_{m}$\\
			\hline 
	\renewcommand{\arraystretch}{1} \\			
			$1.455$& \quad$1.455$& \quad$2.911$&\quad$2.911$& \quad$2.911$ & \quad $2.911$
			\qquad \\		
			\hline
			\end{tabular}
			\label{Tab5}
		\end{table}

	\renewcommand{\arraystretch}{1.5}								
		\begin{table}[b]
			\caption{The numerical values of dimensionless $Q_{\Gamma\gamma}$ and 
			$ R_{m}$ defined in Eqs.~\ref{s3_e23ff}
			and \ref{s3_e25gg} in the classical spin approximation.
			}							
			\center					
			\begin{tabular}{cccccc}
			\hline 
			$Q_{yz}^{\pm}$	&\quad$Q_{zx}^{\pm}$	&\quad$Q_{xy}$	
			&\quad$Q_{2}^{2}$	&\quad$Q_{2}^{0}$ &\quad$R_{m}$\\
			\hline 
	\renewcommand{\arraystretch}{1} \\			
			$-2.121$ & $-2.121$ & $-4.243$ &$-4.243$ & $-4.243$ &  $-4.243$
			 \\		
			\hline
			\end{tabular}
			\label{Tab6}
		\end{table}

\section{} 

The $s$-$f$ interaction is originally described as
		\begin{align}
			-2j({\bm r})\,{\bm s}\cdot{\bm S}.
			\label{eC1}
		\end{align}
Here, $j({\bm r})$ is the exchange integral which is spread out to about the size of $f$ electron.
Assuming free electron approximation,
		\begin{align}
			\int \frac{1}{\sqrt{V}} \exp(i{\bm k}{\bm r})\, j({\bm r})\,\frac{1}{\sqrt{V}} \exp(-i{\bm k'}{\bm r})\,d{\bm r}=j_{0}.
			\label{eC2}
		\end{align}
The relation between $j({\bm r})$ and  $j_{0}$ is as follows.
		\begin{align}
			j({\bm r})=j_{0}V\delta({\bm r}).
			\label{eC3}
		\end{align}

\section{} 
\begin{figure}[t]
		\includegraphics[width=8.5cm]{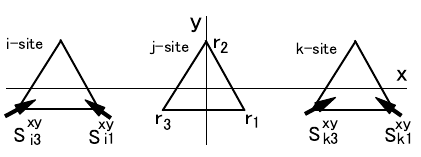}
		\caption{Cancellation of dipole interactions described in Eqs.~\ref{eD4}-\ref{eD6} on the vertices of trimer
		(${\bm r_{1}},{\bm r_{2}},{\bm r_{3}}$) on $j$ site in AFQ phases.  
		The spin polarizations arising from $({\bm S_{k1}^{xy}}-{\bm S_{k3}^{xy}})$ are directed in the opposite
		directions of spin polarizations arising from $({\bm S_{i3}^{xy}}-{\bm S_{i1}^{xy}})$.
		}
		\label{f26}
	\end{figure}

Based on the assumption $R_{0}\gg D,x,y,\Delta r$, we approximated the lengths $l_{i}$ $(i=1,2,3)$ in Fig.~\ref{f10} as follows,
		\begin{align}
			l_{1}&=\left[\left( R_{0}+\Delta r -\frac{D}{2}+x \right)^{2} 
			          +\left(\frac{D}{2\sqrt{3}}+y\right)^{2}\right]^{\frac{1}{2}} \notag\\
			  &\simeq R_{0} \bigg\{ 1 +\frac{1}{2}\bigg[\left(\frac{\Delta r}{R_{0}} \right)^{2} +\left(\frac{D}{2R_{0}}\right)^{2} 
			  + \left(\frac{x}{R_{0}} \right)^{2}  \notag\\
			    &\quad \quad \quad\quad\quad\quad+2\frac{\Delta r}{R_{0}}- \frac{\Delta rD}{R_{0}^{2}} 			      
			       - \frac{Dx}{R_{0}^{2}} -\frac{D}{R_{0}}  \notag \\
			       &\quad\quad \quad \quad\quad\quad\quad+2\frac{\Delta r x}{R_{0}^{2}} +2\frac{x}{R_{0}} 
			       + \left(\frac{y}{R_{0}}\right)^{2} \notag \\
			      &\quad \quad \quad\quad\quad\quad\quad\quad
			       +\frac{1}{12}\left(\frac{D}{R_{0}}\right)^{2}  +\frac{1}{\sqrt{3}}\frac{Dy}{R_{0}^2} \bigg]  \bigg\},
			      \label{eD1}
		\end{align}

		\begin{align}
			l_{2}&=\left[\left( R_{0} +\Delta r +x \right)^{2} +\left(\frac{D}{2\sqrt{3}}-y\right)^{2}\right]^{\frac{1}{2}} \notag\\
			  &\simeq R_{0} \bigg\{ 1 +\frac{1}{2}\bigg[  \left(\frac{\Delta r}{R_{0}} \right)^{2}
			     + \left(\frac{x}{R_{0}} \right)^{2} +2\frac{\Delta r}{R_{0}} \notag \\
			     &\quad\quad\quad\quad\quad\quad\quad +2\frac{\Delta r x}{R_{0}^{2}}
			     +2\frac{x}{R_{0}}  +\left(\frac{y}{R_{0}}\right)^{2} \notag \\
			    &\quad \quad \quad\quad\quad\quad\quad\quad\quad -\frac{2}{\sqrt{3}}\frac{Dy}{R_{0}^{2}}		     
			      +\frac{1}{3}\left(\frac{D}{R_{0}}\right)^{2} \bigg]  \bigg\},
			      \label{eD2}
		\end{align}

		\begin{align}
			l_{3}&=\left[\left( R_{0}+\Delta r +\frac{D}{2}+x \right)^{2} 
				 +\left(\frac{D}{2\sqrt{3}}+y\right)^{2}\right]^{\frac{1}{2}} \notag\\ 
			  &\simeq R_{0} \bigg\{ 1 +\frac{1}{2}\bigg[\left(\frac{\Delta r}{R_{0}} \right)^{2} +\left(\frac{D}{2R_{0}}\right)^{2} 
			  + \left(\frac{x}{R_{0}} \right)^{2}  \notag\\
			    &\quad \quad \quad\quad\quad\quad+2\frac{\Delta r}{R_{0}} + \frac{\Delta rD}{R_{0}^{2}} 			      
			       + \frac{Dx}{R_{0}^{2}} +\frac{D}{R_{0}}  \notag \\
			       &\quad\quad \quad \quad\quad\quad\quad+2\frac{\Delta r x}{R_{0}^{2}} +2\frac{x}{R_{0}} 
			       + \left(\frac{y}{R_{0}}\right)^{2} \notag \\
			      &\quad \quad \quad\quad\quad\quad\quad\quad
			       +\frac{1}{12}\left(\frac{D}{R_{0}}\right)^{2}  +\frac{1}{\sqrt{3}}\frac{Dy}{R_{0}^2} \bigg]  \bigg\}.
			      \label{eD3}
		\end{align}
According to Eq.~\ref{s3_e42gG}- Eq.~\ref{s3_e44bBB}, synthesized spin polarization from 
${\bm S_{1}^{xy}}$, ${\bm S_{2}^{xy}}$, ${\bm S_{3}^{xy}}$,
at ${\bm r_{1}}$ in Fig.~\ref{f10} is given by
		\begin{align}
			{\bm p}({\bm r_{1}})
			&\simeq -C_{p}C_{1} \left[ \frac{D^{2}}{2R_{0}} {\bm S_{i1}^{xy}} -\frac{r_{0}D}{2R_{0}}  ({\bm S_{i3}^{xy}} -  {\bm S_{i1}^{xy}})\right].
			 		\label{eD4}  \\		 
			{\bm p}({\bm r_{2}})
			&\simeq -C_{p}C_{1} \left[\frac{D^{2}}{2R_{0}} {\bm S_{i2}^{xy}} -\frac{r_{0}D}{2R_{0}}  ({\bm S_{i3}^{xy}} -  {\bm S_{i1}^{xy}}) \right].
			 		\label{eD5}	\\	 
			{\bm p}({\bm r_{3}})
			&\simeq -C_{p}C_{1} \left[ \frac{D^{2}}{2R_{0}} {\bm S_{i3}^{xy}} -\frac{r_{0}D}{2R_{0}}  ({\bm S_{i3}^{xy}} -  {\bm S_{i1}^{xy}})\right].
			 		\label{eD6}		 
		\end{align}
Here, $C_{p}$ and $C_{1}$ are both positive. 
In the process of calculation, we are conscious of the relation 
		\begin{align}
			{\bm S_{i1}^{xy}} + {\bm S_{i2}^{xy}} + {\bm S_{i3}^{xy}}={\bm 0}. 
				\label{eD7}
		\end{align}
The terms containing ${\bm S_{i3}^{xy}} -  {\bm S_{i1}^{xy}}$ in Eqs.~\ref{eD4}, \ref{eD5} and \ref{eD6} are canceled in FM or AFM MQ phase due to the interactions from another trimer on the opposite side. As shown in Fig.~\ref{f26},
spin polarization described by ${\bm S_{i3}^{xy}}-{\bm S_{i1}^{xy}}$ and ${\bm S_{k1}^{xy}}-{\bm S_{k3}^{xy}}$ 
are canceled out with each other on the vertices of trimer $j$ site in AFM MQ phase.
Above discussion is valid in the cases of ${\bm S_{i}^{\alpha}}$ $(\alpha=yz, zx)$ also.

\section{} 
According to Eq.~\ref{s2_e3c}, MQ moments are given as
		\begin{align}
			  \tilde{Q}_{\Gamma \gamma}
			=\mu_{0}^{-1}\int \rho_{m} ({\bm r})
				 F_{\Gamma \gamma}({\bm r} ) \, d{\bm r}. 
				\label{eE1}
		\end{align}
When the coordinate $x$--$y$--$z$ is rotated into the coordinate $x'$--$y'$--$z'$ by rotation around 
the $z$ axis $R_{z}(\phi)$,
the $\Gamma_{zx}$ component of $R_{z}\tilde{Q}_{zx}$ is described as
		\begin{align}
			&\mu_{0}^{-1}\int \rho(r)
				\left[ R_{z}^{-1} (z'x') \right]\, zx \, d{\bm r} \notag\\
			&=\mu_{0}^{-1}\int \rho(r)
				\left[  z(x\cos{\phi}+y\sin{\phi}) \right]\, zx \, d{\bm r}\notag\\
			&=\tilde{Q}_{zx}\cos{\phi}. 
				\label{eE4}
		\end{align}
Here, 
		\begin{align}
			\tilde{Q}_{zx}=\mu_{0}^{-1}\int \rho(r)(zx)\, zx \, d{\bm r} .
				\label{eE4}
		\end{align}
Similarly, the $\Gamma_{yz}$ component of $R_{z}\tilde{Q}_{zx}$ is described as
		\begin{align}
			\mu_{0}^{-1}\int \rho(r)
				\left[ R_{z}^{-1} (z'x') \right] yz \, d{\bm r} 
			=\tilde{Q}_{yz}\sin{\phi},
				\label{eE5}
		\end{align}
where
		\begin{align}
			\tilde{Q}_{yz}=\mu_{0}^{-1}\int \rho(r)(yz) \,yz \, d{\bm r} .
				\label{eE4}
		\end{align}

When ${\bm S_{r}}$ is parallel to the $z$ axis,
		\begin{align}
			&\mu_{0}^{-1}\int \rho(r)
				\left[ R_{z}^{-1} \left(\frac{x'^{2}-y'^{2}}{2}\right)\right] \frac{x^{2}-y^{2}}{2} \, d{\bm r} \notag \\
			&=\tilde{Q}_{2}^{2}\cos{2\phi}. 
				\label{eE6}
		\end{align}
Similarly, 
		\begin{align}
			\mu_{0}^{-1}\int \rho(r)
				\left[ R_{z}^{-1} \left( \frac{x'^{2}-y'^{2}}{2}\right)\right] xy \, d{\bm r}
			= \tilde{Q}_{xy}\sin{2\phi}. 
				\label{eF7}
		\end{align}
%

%
%
%
%

\section{} 
\begin{figure}[t]
		\includegraphics[width=8.5cm]{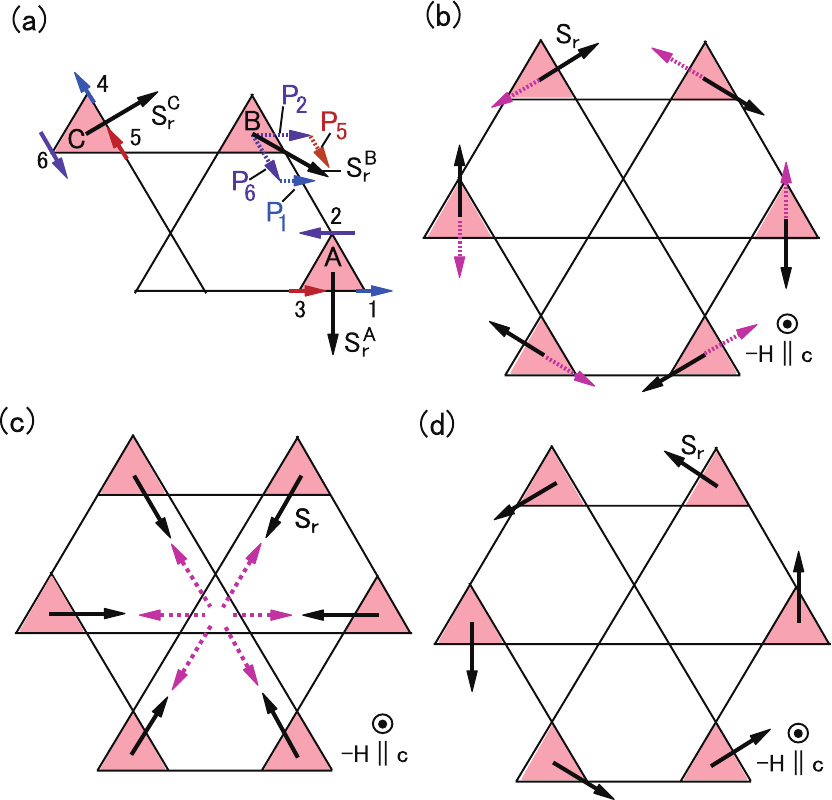}
		\caption{(a) Three type-P trimers A, B and C which are placed at a $120^{\circ}$ angle.
		 The symbols $P_{i}$ ($i=1,2,5,6$) denote spin polarization deduced from the spins on vertices $i$, respectively.
		(b) The circular ${\bm S_{r}}$ structure on type-P TrH expected from Eq.~\ref{s4_e83dDDd}. 
		The solid black
		arrows denote ${\bm S_{r}}$s and the broken purple arrows indicate magnetic field lines arising from ${\bm S_{r}}$s.
		Directions of ${\bm S_{r}}$s are clockwise.
		(c) A radial ${\bm S_{r}}$ structure on type-P hexagon. 
		(d) The circular ${\bm S_{r}}$ structure where directions of ${\bm S_{r}}$s are anti-clockwise.
		}
		\label{f27}
	\end{figure}

Figure~\ref{f27}(a) shows spin structure of three type-P trimers A, B and C which are placed at $120^{\circ}$ angle.
The vectors $P_{i}$ ($i=1,2,5,6$) denote spin polarization deduced from the spins on vertex $i$.
According to Eqs.~\ref{eD4}, \ref{eD5} and \ref{eD6}, ${\bm S_{r}^{A}}$ and  ${\bm S_{r}^{C}}$ produce spin polarization 
 $\sum_{i}{P_{i}}$ on B-site. The synthesized polarization is parallel to ${\bm S_{r}^{B}}$ as shown in Fig.~\ref{f27}(a). 
Therefore, the circular ${\bm S_{r}}$ structure illustrated in Fig.~\ref{f27}(b) is available in Eq.~\ref{s4_e83dDDd}.
The directions of ${\bm S_{r}}$s are clockwise.
The magnetic field lines generated by ${\bm S_{r}}$s are as shown by dotted purple arrows
in Fig.~\ref{f27}(b).
In this case the density of magnetic field lines becomes low in the vicinity of the center of the plaque.
Therefore, electrostatic energy in the vicinity of the center site becomes low, and the structure of ${\bm S_{r}}$ 
in Fig.~\ref{f27}(b) becomes stable.

According to Eq.~\ref{s4_e83dDDd}, the radial spin structure illustrated in Fig.~\ref{f27}(c) is also available .
However, the density of magnetic field lines indicated by dotted purple arrows 
becomes high in the vicinity of the center of the plaque as shown in this figure.
The high density of magnetic field lines increases electrostatic energy in the vicinity of the center site. 
Therefore, the structure shown in Fig.~\ref{f27}(c) is not stable.

The circular ${\bm S_{r}}$ structure is illustrated in Fig.~\ref{f27}(d).
The directions of ${\bm S_{r}}$s are anti-clockwise.
The  electrostatic energy in the vicinity of the center site becomes low similar to Fig.~\ref{f27}(b), 
and the structure of ${\bm S_{r}}$ in Fig.~\ref{f27}(d) is stable. 
The structure of ${\bm S_{r}}$ in Fig.~\ref{f27}(b) and Fig.~\ref{f27}(d) are replaced with each other by the inversion operation
on the $a$ plane. Therefore, the circular structures in Figs.~\ref{f27}(b) and \ref{f27}(d) have ${\bm Z_{2}}$ mirror symmetry.
This suggests that clockwise and anti-clockwise domains are generated in SkL phase. These structures are topologically
stable because of the high energy of the structure in Fig.~\ref{f27}(c)
\cite{Everschor-Sitte2018}. In the SkL phase,  ${\bm Z_{2}}$ mirror symmetry is broken.

	\begin{figure}[b]
		\includegraphics[width=8.5cm]{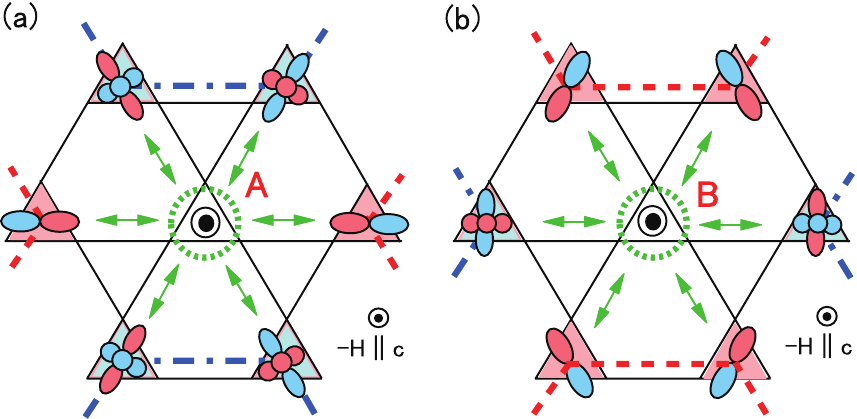}
		\caption{(a) The arrangement of QSs around (a) site ``A'' and (b) around site ``B''
		in Fig.~\ref{f23} on expanded scales. The shapes around the center sites indicate project drawings of QSs
		shown in Figs.~\ref{f21} and \ref{f22}.
		The bi-directional green arrows indicate MQ interactions. The dotted green circles denotes quadrupole disorder.
		}
		\label{f28}
	\end{figure}

\section{} 

Quadrupole disorder realize other than the center sites of TrHs illustrated in Fig.~\ref{f23}. Figure \ref{f28}(a)
represents the QS arrangements around the site ``A'' in Fig.~\ref{f23} on expanded scales. The bi-directional green arrows
in this figure denote the MQ interactions. The shapes on the trimers around A-site indicates the project drawings
of QSs shown in Figs.~\ref{f21} and \ref{f22} along the $c$ axis.
As shown in Fig.~\ref{f28}(a), MQ interactions from these six surrounding trimer sites cancel out with 
each other on A-site, therefore,
site-A becomes to be a quadrupole disorder site. The dotted green circle on A-site denotes quadrupole disorder.
The arrangement of QSs around the site-B in Fig.~\ref{f23} is displayed in Figure \ref{f28}(b) on expanded scales.
Similar to the case of Fig.~\ref{f28}(a), MQ interactions are canceled out each other on B-site, and this site
becomes to be a quadrupole disordered site. Hence, the lattice of TrHs enfolds many quadrupole disordered sites
in it under fields directed along the $c$ axis. The coexistence of TrHs with different anisotropies results
in many trimer sites with quadrupole disorder.


\end{document}